\documentclass[10pt,journal,twoside,final]{IEEEtran}

\newtheorem{theo}{Theorem}
\newtheorem{lemma}{Lemma}

\IEEEoverridecommandlockouts

\makeatletter
\newcommand{\biggg}[1]{{\hbox{$\left#1\vbox to 20.5pt{}\right.\n@space$}}}
\newcommand{\Biggg}[1]{{\hbox{$\left#1\vbox to 23.5pt{}\right.\n@space$}}}
\newcommand{\bigggg}[1]{{\hbox{$\left#1\vbox to 26.5pt{}\right.\n@space$}}}
\newcommand{\Bigggg}[1]{{\hbox{$\left#1\vbox to 29.5pt{}\right.\n@space$}}}
\newcommand{\biggggg}[1]{{\hbox{$\left#1\vbox to 32.5pt{}\right.\n@space$}}}
\newcommand{\Biggggg}[1]{{\hbox{$\left#1\vbox to 35.5pt{}\right.\n@space$}}}
\newcommand{\bigggggg}[1]{{\hbox{$\left#1\vbox to 38.5pt{}\right.\n@space$}}}
\newcommand{\Bigggggg}[1]{{\hbox{$\left#1\vbox to 41.5pt{}\right.\n@space$}}}
\makeatother

\usepackage{bm,cite,algorithm,algorithmic,float,amsmath,amssymb}
\usepackage[dvips]{graphicx}
\usepackage{subfigure,amsfonts}
\usepackage{mdwmath}
\usepackage{mdwtab}
\usepackage{xcolor,color}
\usepackage{cool}
\usepackage{balance}
\usepackage{diagbox}
\floatname{algorithm}{Algorithm}

\makeatletter
\renewcommand\paragraph{\@startsection{paragraph}{4}{\z@}%
            {-2.5ex\@plus -1ex \@minus -.25ex}%
            {1.25ex \@plus .25ex}%
            {\normalfont\normalsize\itshape}}
\makeatother
\usepackage{epstopdf}

\usepackage[style=0]{mdframed}
\usepackage{showexpl}

\mdfdefinestyle{exampledefault}{%
rightline=true,innerleftmargin=10,innerrightmargin=10,
frametitlerule=true,frametitlerulecolor=green,
frametitlebackgroundcolor=yellow,
frametitlerulewidth=2pt}

\allowdisplaybreaks

\begin{document}

%\markboth{IEEE Transactions on Wireless Communications,~Vol.~X,
%	No.~XX,~2020}{Liu \MakeLowercase{\textit{et al.}} : Rate Splitting for Uplink NOMA with Enhanced Fairness and Outage Performance}

\title{Covert Communications in STAR-RIS-Aided Rate-Splitting Multiple Access Systems}

\author{
Heng Chang, Hai Yang, Shuobo Xu, Xiyu Pang, and Hongwu~Liu 

\thanks{H. Chang, H. Yang, S. Xu, X. Pang, and H. Liu are with the School of Information Science and Electrical Engineering, Shandong Jiaotong University, Jinan 250357, China (emails: 21208034@stu.sdjtu.edu.cn, yh\_sdjtu@163.com, shuobo@163.com, xiyupang@126.com, liuhongwu@sdjtu.edu.cn). }

}

\maketitle
\setcounter{page}{1}
\begin{abstract}
In this paper, we investigate covert communications in a simultaneously transmitting and reflecting reconfigurable intelligent surface (STAR-RIS)-aided rate-splitting multiple access (RSMA) system. Under the RSMA principles, the messages for the covert user (Bob) and public user (Grace) are converted to the common and private streams at the legitimate transmitter (Alice) to realize downlink transmissions, while the STAR-RIS is deployed not only to aid the public transmissions from Alice to Grace, but also to shield the covert transmissions from Alice to Bob against the warden (Willie). To characterize the covert performance of the considered STAR-RIS-aided RSMA (STAR-RIS-RSMA) system, we derive analytical expression for the minimum average detection error probability of Willie, based on which a covert rate maximization problem is formulated. To maximize Bob's covert rate while confusing Willie's monitoring, the transmit power allocation, common rate allocation, and STAR-RIS reflection/transmission beamforming are jointly optimized subject to Grace's quality of service (QoS) requirements. The non-convex covert rate maximization problem, consisting of highly coupled system parameters are decoupled into three sub-problems of transmit power allocation, common rate allocation, and STAR-RIS reflection/transmission beamforming, respectively. 
To obtain the rank-one constrained optimal solution for the sub-problem of optimizing the STAR-RIS reflection/transmission beamforming, a penalty-based successive convex approximation scheme is developed. Moreover, an alternative optimization (AO) algorithm is designed to determine the optimal solution for the sub-problem of optimizing the transmit power allocation, while the original problem is overall solved by a new AO algorithm. Simulation results corroborate the accuracy of the derived analytical results and demonstrate that the proposed STAR-RIS-RSMA scheme achieves a higher covert rate than the benchmark schemes.
\end{abstract}


\section{Introduction}

As a novel multiple access technology, rate-splitting multiple access (RSMA) attracted great attention due to the advantages of providing flexible interference management, achieving massive connectivity and optimal user-fairness, and improving energy efficiency and spectrum efficiency   \cite{9598915,9461768,Sum_rate_RS_approach,RS_unifying, RSMA_new_frontier,RS_EURASIP,RSMA_Survey_Trends}. In RSMA systems, each user's message is split into the common and private parts, based on which the common and private streams are, respectively, generated to realize the downlink transmissions. At the receiver side, the dynamic interference management strategy and successive interference cancellation (SIC) are utilized to recover the desired signal. As those conducted in the one-layer RSMA, each user decodes the common stream in the first stage of SIC by treating all the private streams as noise \cite{RS_EURASIP}. Then, each user decodes its own private stream in the second stage of the SIC by treating the other users' private streams as noise, which achieves a promising system performance while maintaining a moderate decoding complexity \cite{RS_EURASIP}. Till now, exploring the potential performance enhancements provided by RSMA, a new start of research trends on RSMA under both overloaded and underloaded scenarios is expected \cite{RSMA_Survey_Trends}.

In the upcoming sixth-generation (6G) wireless communications, the volume of confidential information and corresponding sensitive transmissions tend to explode dramatically, which results in serious security issues on the developments of radio access technologies. From the perspective of the information theory, physical layer security (PLS) can prevent eavesdroppers from correctly decode the legitimate information by utilizing various transmission techniques, e.g., transmit power allocation, beamforming, and artificial noise (AN) \cite{PLS_Survey}. In \cite{9217123}, the common message was used to not only convey information for multiple RSMA users, but also act as AN for interfering the eavesdroppers. Considering imperfect channel state information at transmitter (CSIT), the common message was also exploited as jamming to confuse the eavesdropper such that the secrecy sum-rate of the RSMA users can be maximized \cite{9195771}. In the presence of an internal eavesdropper, a secure beamforming was proposed to enhance the secrecy-rate performance subject to all users' secrecy rate constraints \cite{9771854}. Moreover, a joint optimization of the precoding, slot allocation, and rate allocation was proposed to safeguard the legitimate information transmissions in the RSMA against the untrusted relays \cite{9910034}. 
Compared to PLS, covert communications provided a higher level of security by hiding the legitimate communication behaviors against adversaries, i.e., rending a large probability of incorrect detection for wardens' monitoring such that 
wardens cannot be correctly aware of the legitimate communication behaviors \cite{8883125, 6584948}. In military and sensitive civilian applications, exposing transmission behaviors or leaking a small amount of data can result in unexpected detrimental impacts to legitimate systems, which generates unprecedented demanding for covert communications and becomes an emerging challenge for the RSMA systems. In \cite{10038571}, the joint optimization of the transmit power and rate allocations was investigated and a gradient-based learning algorithm was proposed to maximize the sum rate and maintain fairness among RSMA users. Furthermore, by superimposing AN to the common and private messages, a jamming strategy was proposed to prevent the warden from being aware of the legitimate RSMA transmissions while the minimum data rate among RSMA users could be significantly increased \cite{10080970}. 

On the other hand, reconfigurable intelligent surfaces (RISs) have the capability of configuring the radio environment \cite{8796365, 8811733, 9140329, SCRIS2, SCRIS1}. By deploying a large number of low-cost reflecting elements, a RIS can intelligently adjust the phase shifts and/or amplitudes of reflected signals, which introduces, but not limited to coverage enhancements, high spectral efficiency, high energy efficiency, and reliable transmissions for wireless communications. Empowered by RISs, covert transmissions can be safeguarded by changing wireless propagation environments and a joint configuration of covert transmission and RIS's reflection can achieve a considerable improvement on covert performance \cite{9108996}. Based on either the full or partial channel state information (CSI) of adversary link, the coupled transmit power and RIS phase shifts were alternatively optimized to maximize the covert rate \cite{9438645}. The work in \cite{9496108} characterized CSI conditions to achieve perfect covertness in RIS-aided covert communication systems in the cases with global CSI and without instantaneous adversary CSI. In \cite{9363936}, the joint design of active and passive beamformings was investigated to maximize the covert rate of RIS-aided systems, where a multi-antenna transmitter and a legitimate full-duplex receiver conducted covert communications in the presence of a watchful warden. With the availability of statistical adversary CSI, both reflection phase shifts and amplitudes were jointly optimized to enhance covert performance \cite{9390203}. Moreover, RISs were applied in non-orthogonal multiple access (NOMA) systems to assist covert transmissions from a legitimate transmitter to NOMA users. In \cite{9524501}, transmit power and phase shifts were jointly optimized to increase covert rate subject to the given covertness against warden, and the effects of RIS-aided NOMA transmissions on covert performance were revealed. In RIS-aided NOMA system, a jammer was deployed to hide to the existence of strong user, which further improved covert performance \cite{10001489}. Nevertheless, employing RIS to enhance covertness of future multiple access systems is still in infancy.

Although users located on the reflection side of the conventional RISs can receive the reflected signals to enhance wireless network performance, users located behind cannot be served due to geographical limitations.
To break this bottleneck of the conventional RISs, simultaneous transmitting and reflecting RIS (STAR-RIS) has been proposed to provide a $360^{\circ}$ full coverage \cite{9690478,9437234,9365009}.
In \cite{9690478}, the basic physical layer modes, including energy splitting (ES), mode switching (MS), and time switching (TS), have recently been proposed for STAR-RISs. Since STAR-RISs provides full-space service coverage, NOMA users located on the reflection and transmission sides of the STAR-RIS can be simultaneously connected to the same base-station (BS). Although combing STAR-RIS and NOMA is a win-win strategy to improve coverage performance, eavesdroppers may enjoy similar performance gains as legitimate users. In \cite{9915477} and \cite{9834288}, how to exploit STAR-RISs to enhance secrecy performance was investigated for downlink and uplink NOMA systems, respectively. For the STAR-RIS-aided multiple-input and single-output (MISO) system, the authors in \cite{9525400} investigated the secrecy  performance under three transmission modes, i.e., ES, MS, and TS, and jointly designed   transmit beamforming and RIS reflection and transmission coefficients to maximize the weighted secrecy sum-rate. Moreover, AN was applied to improve secrecy performance of STAR-RIS-aided NOMA (STAR-RIS-NOMA) systems \cite{9739715}.
Very recently, STAR-RISs have been applied in RSMA systems to enhance system performance.
In \cite{9854887}, outage probability and channel capacity of the STAR-RIS-aided downlink RSMA system have been analyzed for both infinite and finite transmission block lengths. In \cite{10014691}, the STAR-RIS has been deployed to improve the spectral efficiency of the uplink RSMA system. However, how to exploit STAR-RISs to safeguard PLS of RSMA system is challenging. From privacy protections perspective, the early stage researches on STAR-RIS-aided RSMA (STAR-RIS-RSMA) covert communications are urgently demanded.  

\subsection{Motivation and Our Contributions}

As outlined above, existing works on security and privacy protection of RIS-aided wireless systems mainly focused on the conventional RIS mode (reflection). Since STAR-RISs introduce transmission coefficients in addition to reflection ones, these existing results cannot be applied to STAR-RIS-aided wireless systems. Furthermore, current works \cite{9915477,9834288,9525400,9739715} on STAR-RIS-aided PLS are confined to NOMA systems, which, however, are non-trivial to be implemented in general covert RSMA systems. This is due to fact that STAR-RIS reflection and transmission coefficients are highly coupled with common and private messages, which is uniquely belonged to RSMA in comparison with NOMA. Therefore, it is important to consider STAR-RIS-aided covert communications in RSMA systems. To realize covert communications between the legitimate transmitter and users, the common and private messages, transmit power, and STAR-RIS reflection and transmission coefficients need to be considered. To fully exploit the uncertainties of common and private messages, transmit power, and STAR-RIS reflection and transmission coefficients to realize covert communications, we consider a STAR-RIS-RSMA system and analyze the corresponding detection error probability (DEP) as a covert metric. Then, we conduct the optimal covert communication design for the considered STAR-RIS-RSMA system.

The main contributions of this paper are summarized as follows:

\begin{itemize}
\item We investigate covert communications in a STAR-RIS-RSMA system, where a warden and a covert user can receive signals reflected by the STAR-RIS, and transmitted by a legitimate user via the direct links. The transmit power, common and private messages, and STAR-RIS reflection/transmission coefficients are jointly optimized as new uncertainty medium to protect the private message transmissions of the covert user.  
\item For the worst-case scenario of the legitimate system, also the most favorable case to warden, we derive a closed-form expression for the minimum average detection error probability (MADEP) by assuming that warden can optimally choose its detection threshold. Moreover, MADEP characterizes the covertness performance of the considered STAR-RIS-RSMA system, and provides a covert metric for the optimal design of covert communications. 
\item Subject to the QoS requirements of the public user and the covertness constraint acquired by the system, the transmit power allocation, rate allocation, and STAR-RIS reflection/transmission beamforming are jointly optimized to maximize covert rate. To solve the non-convex covert rate maximization problem, an efficient alternating optimization (AO) algorithm to dealing with three sub-problems of optimizing the transmit power allocation, rate allocation, and STAR-RIS reflection/transmission coefficients, respectively, is proposed. A penalty-based successive convex approximation (PSCA) method is proposed to obtain the rank-one optimal solution to the sub-problem optimization of the STAR-RIS coefficients.  
\end{itemize}

The rest of the paper is organized as follows. 
The system model of the STAR-RIS-RSMA system is presented in Section 2. In Section 3, the covert performance of the considered system is analyzed. Section 4 formulates the covert rate maximization problem and develops the optimization method. Simulation results are presented in Section 5 and concluding remarks are given in Section 6.

$Notations$: Bold lowercase and uppercase letters represent vectors and matrices, respectively. $\mathbb{E}(\cdot)$ represent the expectation. $\rm{diag}(\cdot)$ constructs a diagonal matrix from its vector argument. $\mathbb{C}^{N \times M}$ stands for the set of $N \times M$ complex matrices. $ {\bf{h}}_i \sim \mathcal{CN}({\bf{0}}, \lambda_i {\bf{I}})$ denotes that ${\bf{h}}_i$ is the complex Gaussian random variable with zero mean and variance is $\lambda_i {\bf{I}}$, where ${\bf{I}}$ is a all ones vector with a proper length. $\Pr(\cdot)$ denotes probability. The operators $\rm{Tr}(\cdot)$ and $\rm{rank}(\cdot)$ denote the trace and rank of a matrix, respectively.

\section{System Model}
We consider a STAR-RIS-RSMA system as shown in Fig. 1, in which a $K$-element STAR-RIS is deployed to assist the transmissions from a transmitter (Alice) to a covert user (Bob) and a public user (Grace) by using downlink RSMA. Without loss of generality, we assume that Bob and Grace are near and far users, respectively, with their locations in the reflection and transmission sides of the STAR-RIS, respectively, while a warden (Willie) operates within the vicinity of Bob to monitor the covert transmissions from Alice to Bob. Due to severe fading and blocking, we assume the direct link between Alice and Grace is unavailable in this study, while Bob and Willie can receive signals transmitted from Alice directly and reflected by the STAR-RIS.

In this study, we have the following assumptions for the considered STAR-RIS-RSMA system: 1) All nodes are equipped with a single antenna for reception or transmission except the STAR-RIS consisting of $K$ elements. 2) STAR-RIS has three operating protocols. For the ES mode, all elements are designed with a high degree of flexibility, however, there will be a high overhead in designing the exchange of configuration information from Alice to STAR-RIS to the user. For the TS mode, reflective and transmission elements are easier to design due to the utilization of the time domain, however the periodic switching of the elements places extreme demands on the hardware. For the MS mode, the reflective and transmission elements can be designed independently of each other \cite{9570143}. From a practical point of view, such an "on/off" operating protocol of the MS mode can reduces the difficulty of optimizing the variables for subsequent design \cite{9690478}. Therefore, we assume that STAR-RIS works in MS mode, i.e., all the $K$ elements are divided into two groups with $K_n$ elements in one group working in the reflection mode to serve Bob and the remaining $K_m$ ($K_m = K - K_n$) elements in another group working in the transmission mode to serve Grace. 3) By using channel estimation method in \cite{9039554,8937491}, Alice knows the instantaneous CSI of the direct-link from Alice to Bob, the reflecting channels from Alice via the STAR-RIS to Bob, and the refracting channels from Alice through the STAR-RIS to Grace.  4) Alice knows the statistical CSI of the channels from Alice to Willie, which can be obtained by observing the location information of Willie, whereas Alice does not know the corresponding instantaneous CSI because Willie tries to hide its existence from the legitimate system. 5) From the perspective of covert communication design, we consider the worst-case scenario, i.e., Willie knows the instantaneous CSI of the direct-link from  Alice to Willie and the reflecting channels from Alice through the STAR-RIS to Willie by estimating the pilots sent by Alice \cite{9524501}. 6) All channels are modeled by a small-scale fading distribution along with a distance-based path-loss. The small-scale fading coefficients associated with the STAR-RIS are modeled by the vector $ {\bf{h}}_i \sim \mathcal{CN}({\bf{0}}, \lambda_i {\bf{I}})$ with a corresponding dimension ($K_n \times 1$ or $K_m \times 1$). Moreover, the subscript of $ {\bf{h}}_i$ is set by $i \in \{ar_1, ar_2, r_1b, r_2g, r_1w \}$ with $a$, $b$, $g$, $w$, $r_1$, and $r_2$ stand for Alice, Bob, Grace, Willie, the group of $K_n$ STAR-RIS reflection elements, and the group of $K_m$ STAR-RIS transmission elements, respectively. For ease of notations, we denote ${\bf{h}}_{r_1 b}$ by ${\bf{h}}_{rb}$, ${\bf{h}}_{r_2 g}$ by ${\bf{h}}_{rg}$, and ${\bf{h}}_{r_1 w}$ by ${\bf{h}}_{rw}$ and set  $\lambda_{ar_1} = \lambda_{ar_2} = \lambda_{ar}$ considering that the two groups of STAR-RIS elements are located on the same base plane.
The small-scale fading coefficients of the direct-link channels from Alice to Bob and Willie are modeled by $h_i\sim \mathcal{CN}(0, \lambda_i)$ with $i\in\{ab, aw\}$. For the channel $h_i$ (or ${\bf h}_i$), the associated path-loss is denoted by $L_{i} = L_0 \big(\frac{d_i}{d_0}\big)^{\chi_i}$, where $\chi_i$ is the path-loss factor, $d_i$ is the distance between the two nodes, $d_0$ is the reference distance, and $L_0$ is the path-loss at the reference distance.

\begin{figure}[t]
    \begin{center}
    \includegraphics[width=2.7in]{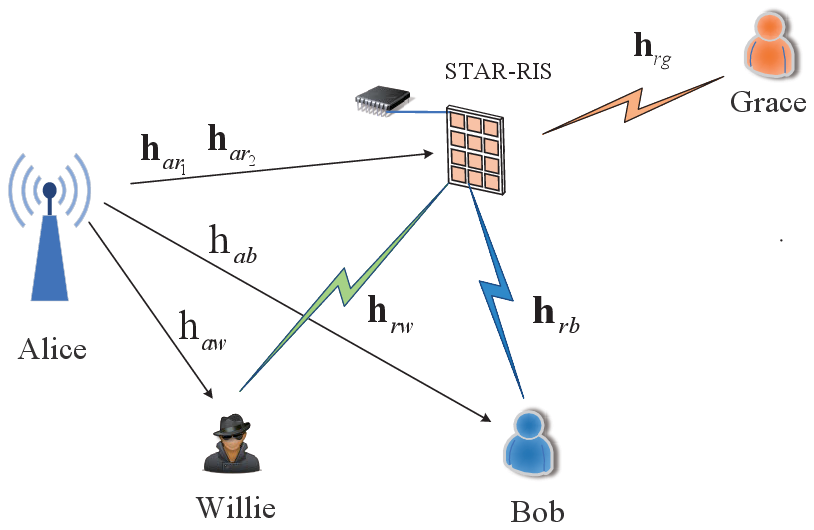}
    \caption{A STAR-RIS-RSMA system conducting covert communications}
    \label{fig:subfig1}
    \end{center}
    \vspace{-0.15in}
\end{figure}

\subsection{RSMA Transmissions}

In the considered STAR-RIS-RSMA  system, Bob receives signals from both the direct link between Alice and Bob and the reflecting link from Alice through STAR-RIS to Bob, while Grace can only receive signals from Alice via the STAR-RIS transmissions. Let $\tilde s_b(q)$ and $\tilde s_g(q)$ denote the messages to be transmitted to Bob and Grace, respectively, where $q = 1,\dots, Q$ is the index of the signal samples and $Q$ refers to the total number of signal samples in a communication slot. 
Before the start of each transmission block, Alice splits $\tilde s_b(q)$ and $\tilde s_g(q)$ into $\{\tilde s_{b}^{(1)}(q), \tilde s_{b}^{(2)}(q)\}$ and $\{\tilde s_{g}^{(1)}(q), \tilde s_{g}^{(2)}(q)\}$, respectively. Then, messages $\tilde s_{b}^{(1)}(q)$ and $\tilde s_{g}^{(1)}$ are merged and encoded to the common stream $s_0(q)$, while $\tilde s_{b}^{(2)}(q)$ and $\tilde s_{g}^{(2)}(q)$ are encoded to the private streams $s_1(q)$ and $s_2(q)$, respectively, which are intended to be transmitted to Bob and Grace, respectively. At the end of each transmission block, the signals received by Bob and Grace can be expressed as:
\begin{eqnarray}
y_{b}(q) = \left(\frac{h_{ab}}{\sqrt{L_{ab}}} + \frac{{\bf{h}}_{ar_{1}}^{H} {\bf{\Theta}}_r{\bf{h}}_{rb}}{\sqrt{L_{ar} L_{rb}}} \right) x_1 + n_b (q)
\end{eqnarray}
and
\begin{eqnarray}
y_{g}(q) =  \frac{{\bf{h}}_{ar_{2}}^{H} {\bf{\Theta}}_t{\bf{h}}_{rg}}{\sqrt{L_{ar} L_{rg}}}  x_1 + n_g (q),
\end{eqnarray}
respectively, where $x_1 = \left( \sqrt{\alpha_0 P_T} s_0(q) + \sqrt{\alpha_1 P_T} s_1(q) + \sqrt{\alpha_2 P_T} s_2(q) \right)$, $P_T$ is Alice's maximum transmitting power, $\alpha_0$, $\alpha_1$, and $\alpha_2$ are transmit power allocation factors
satisfying $0 \le \alpha_0, \alpha_1, \alpha_2 \le 1$, and $n_b (q) \sim \mathcal{CN} (0, \sigma_{b}^{2})$ and $n_g (q) \backsim \mathcal{CN} (0, \sigma_{g}^{2})$ are the additive white Gaussian noise (AWGN) at Bob and Grace, respectively. Moreover,  ${\bf{\Theta}}_r = {\rm diag} \big([e^{j\theta_{1}^{r}}, \cdots, e^{j\theta_{k}^{r}} , \cdots, e^{j\theta_{K_n}^{r}}]\big) $ and ${\bf{\Theta}}_t = {\rm diag} \big([e^{j\theta_{1}^{t}}, \cdots, e^{j\theta_{k}^{t}}, \cdots, e^{j\theta_{K_m}^{t}}]\big) $ are the diagonal matrices with their diagonal elements denoting the STAR-RIS reflection and transmission coefficients, respectively, and $\theta_{k}^{r} \in [0, 2\pi)$ ($\theta_{k}^{t} \in [0, 2\pi)$) denotes the reflection phase shift (transmission phase shift) of the $k$th element of the STAR-RIS reflection group (transmission group). According to the one-layer RSMA principles  \cite{RS_EURASIP}, Bob and Grace treat all the private streams as noise in the detection of the common stream $s_0(q)$. Therefore, the achievable rates obtained by Bob and Grace from detecting $s_0 (q)$ can be expressed as:
\begin{eqnarray}
R_{b,s_0 (q)} = \log_2 \left(1 + \frac{\alpha_0 P_T |Z_{ab}|^2}{(\alpha_1 + \alpha_2)P_T|Z_{ab}|^2 + \sigma_{b}^{2}}\right),
\end{eqnarray}
and 
\begin{eqnarray}
R_{g,s_0 (q)} = \log_2 \left(1 + \frac{\alpha_0 P_T |Z_{ag}|^2}{(\alpha_1 + \alpha_2)P_T|Z_{ag}|^2 + \sigma_{g}^{2}}\right)
\end{eqnarray}
respectively.
where $Z_{ab} \triangleq \tfrac{h_{ab}}{\sqrt{L_{ab}}} + \tfrac{{\bf{h}}_{ar_{1}}^{H} {\bf{\Theta}}_r{\bf{h}}_{rb}}{\sqrt{L_{ar} L_{rb}}}$ and $Z_{ag} \triangleq  \tfrac{{\bf{h}}_{ar_{1}}^{H} {\bf{\Theta}}_t{\bf{h}}_{rg}}{\sqrt{L_{ar} L_{rg}}}$ denotes the composite channel from Alice to Bob and Alice to Grace, respectively. After successfully detecting $s_0$ and using the SIC processing, Bob and Grace treat each other's private stream as noise in the detecting of their own private streams, respectively, so that the achievable rates obtained by Bob and Grace  from detecting their own private streams can be expressed as:
\begin{eqnarray}
R_{b,s_{1} (q)} = \log_2 \left(1 + \frac{\alpha_1 P_T |Z_{ab}|^2}{\alpha_2P_T|Z_{ab}|^2+ \sigma_{b}^{2}}\right)
\end{eqnarray}
and
\begin{eqnarray}
R_{g,s_{2} (q)} = \log_2 \left(1 + \frac{a_2 P_T |Z_{ag}|^2}{\alpha_1 P_T|Z_{ag}|^2 + \sigma_{g}^{2}}\right),
\end{eqnarray}
respectively.

In the considered STAR-RIS-RSMA system, the strengths of the channel gains mainly depend on the direct links due to the severe 'double fading' effect of the STAR-RIS. Since Grace is located at the transmission end of the STAR-RIS and the direct link between Alice and Grace is severely blocked, the composite channel gains satisfy  $|Z_{ag}|^{2} < |Z_{ab}|^2$, i.e., the composite channel gain of Grace is smaller than that of the Bob  \cite{9524501,9146329}. Consequently, we have $R_{g, s_0 (q)} < R_{b, s_0 (q)}$. To ensure that both Bob and Grace can successfully decode the common stream $s_0$, according to the one-layer RSMA principles \cite{RS_EURASIP,9461768}, the common rate can not exceed
\begin{eqnarray}
  R_c &=& \min\big\{R_{b, s_0 (q)}, R_{g, s_0 (q)} \big\}.
\end{eqnarray}
Considering that $R_c$ is the total achievable rate corresponding to the transmissions of the sub-messages $\tilde s_{b}^{(1)}(q)$ and $\tilde s_{g}^{(1)}(q)$, we let $\beta_1$ and $\beta_2$ denote the common rate allocation factors for Bob and Grace, respectively, where $0\le \beta_1,\beta_2 \le 1$ and $\beta_1 + \beta_2 \le 1$. Then, the common rates obtained by Bob and Grace are determined by $R_{b}^{c} = \beta_1 R_{g,s_0 (q)}$ and $R_{g}^{c} = \beta_2 R_{g,s_0 (q)}$, respectively, satisfying $R_{b}^{c} + R_{g}^{c} = R_c$. Consequently, the available rates of Bob and Grace can be given as $R_b = \beta_1 R_{g,s_0 (q)} + R_{b,s_1(q)}$ and $R_g = \beta_2 R_{g,s_0 (q)} + R_{g,s_2(q)}$, respectively.

\subsection{Willie's Detection}

With the received signal, Willie needs to make a judgment on whether Alice transmits a private message to Bob. Let $\mathcal{H}_0$ denote the null hypothesis that Alice does not transmit a private stream to Bob and $\mathcal{H}_1$ denote the alternative hypothesis that Alice transmits a private stream to Bob. Corresponding to $\mathcal{H}_0$ and $\mathcal{H}_1$, the signal received at Willie can be written as:
\begin{eqnarray}
\mathcal{H}_0 : ~~y_{w} (q) = \frac{h_{aw}}{\sqrt{L_{aw}}} x_0 + \frac{{\bf{h}}_{ar_{1}}^{H} {\bf{\Theta}}_r{\bf{h}}_{rw}}{\sqrt{L_{ar} L_{rw}}}  x_0 + n_w \label{H0}
\end{eqnarray}
and
\begin{eqnarray}
\mathcal{H}_1 : ~~y_{w} (q) = \frac{h_{aw}}{\sqrt{L_{aw}}} x_1 + \frac{{\bf{h}}_{ar_{1}}^{H} {\bf{\Theta}}_r{\bf{h}}_{rw}}{\sqrt{L_{ar} L_{rw}}} x_1 + n_w , \label{H1}
\end{eqnarray}
respectively, where $x_0 = \left(\sqrt{\alpha_0 P_T}s_0 (q) + \sqrt{\alpha_2 P_T}s_2 (q)\right)$ is Alice's transmitted signal in the absence of the covert transmissions and $n_w (q) \backsim \mathcal{CN} (0, \sigma_{w}^{2})$ denotes the AWGN at Willie. Based on \eqref{H0} and \eqref{H1}, Willie adopts a radiometer to realize the binary decision and uses the average received power,  $P_w = \frac{1}{Q} \sum_{q=1}^{Q} |y_w(q)|^2$, as the test statistic. To eliminate the uncertainties of the transmitted signal and AWGN, the assumption $Q \to \infty$ is adopted in the performance analyses and system designs \cite{9524501,8654724,9363936}. Therefore, the average received power at Willie can be asymptotically approximated as \cite{9524501,8654724,9363936}:
\begin{small}
\begin{align}
        P_w =~~~~~~~~~~~~~~~~~~~~~~~~~~~~~~~~~~~~~~~~~~~~~~~~~~~~~~~~~~~~~~~~~~~~~~~~ \nonumber \\ \left\{  {\begin{array}{*{20}{c}}
            {\tfrac{|h_{aw}|^2}{L_{aw}} (\alpha_0 \!+\! \alpha_2)P_T + \tfrac{|{\bf{h}}_{ar_{1}}^{H} {\bf{\Theta}}_r{\bf{h}}_{rw}|^2}{L_{ar} L_{rw}} (\alpha_0 \!+\! \alpha_2)P_T + \sigma_{w}^{2} }, & \!\!{ \mathcal{H}_0},\\
            \tfrac{|h_{aw}|^2}{L_{aw}} P_T + \tfrac{|{\bf{h}}_{ar_{1}}^{H} {\bf{\Theta}}_r{\bf{h}}_{rw}|^2}{L_{ar} L_{rw}} P_T + \sigma_{w}^{2}  ,~~~~~~~~~~~ & { \mathcal{H}_1}.
            \end{array}} \right. \!\! \label{P_w}
\end{align}
\end{small}
Based on \eqref{P_w}, Willie adopts the binary decision to detect whether the covert transmission occurs or not and the decision rule can be expressed as:
\begin{eqnarray}
P_w \mathop{\gtrless}\limits_{\mathcal{D}_0 }^{\mathcal{D}_1} \eta, \label{P1_w}
\end{eqnarray}
where $\mathcal{D}_0$ and $\mathcal{D}_1$ are the binary decisions in favor of $\mathcal{H}_0$ and $\mathcal{H}_1$, respectively, and $\eta > 0$ is the detection threshold.

\section{Covert Performance Analysis}

Following the binary decision rule, Willie may make an error decision under the two hypotheses: 1) The covert transmissions to Bob are detected by Willie under the hypothesis $\mathcal{H}_0$, which can be measured by the false alarm probability $\mathbb{P}_{\rm FA} = \Pr(\mathcal{D}_1|\mathcal{H}_0) $. 2) The covert transmissions to Bob are not detected by Willie under the hypothesis $\mathcal{H}_1$, which can be measured by the missed detection probability $\mathbb{P}_{\rm MD} = \Pr(\mathcal{D}_0|\mathcal{H}_1) $. Assuming $\mathcal{H}_0$ and  $\mathcal{H}_1$ have an equal prior probability, the  DEP at Willie can be expressed as $\xi_{w,\eta} = \mathbb{P}_{\rm FA} + \mathbb{P}_{\rm MD}$, which satisfies $\xi_{w,\eta} \in [0,1]$ with $\xi_{w,\eta} = 0$ indicating that Willie can always accurately detect the covert transmissions from Alice to Bob, while $\xi_{w,\eta} = 1$ denoting that Willie cannot detect any covert transmissions from Alice to Bob.

{\it{\textbf{Remark~1:}}} Based on CSI of the Alice-Bob and Alice-STAR-IRS-Bob links, an optimal ${\bf{\Theta}}_r$ can be designed such that the cascaded channel gain  $|{\bf{h}}_{ar_{1}}^{H} {\bf{\Theta}}_r{\bf{h}}_{rw}|^2$ becomes a random variable for Willie, while Willie can only acquire the distribution of $|{\bf{h}}_{ar_{1}}^{H} {\bf{\Theta}}_r{\bf{h}}_{rw}|^2$. This is because when Willie makes the decision based on its received power, it cannot ensure whether a variation in the received power is resulted by the public transmission from Alice to Grace or the covert transmission from Alice to Bob.

For ease of notations, let $H_N \triangleq {\bf{h}}_{ar_{1}}^{H} {\bf{\Theta}}_r{\bf{h}}_{rw}$, which can be approximated as a complex Gaussian random variable with zero mean and variance $\lambda_N  = \lambda_{ar} \lambda_{rw} K_n$ \cite{9079918,9524501}.
Based on \eqref{P_w} and \eqref{P1_w}, the DEP of Willie in the considered STAR-RIS-RSMA system with an arbitrary detection threshold can be derived as:
\begin{align}
\xi_{w,\eta} = \Pr \bigg(\frac{(\alpha_0 +\alpha_2)P_T |h_{aw}|^2}{L_{aw}} + \frac{(\alpha_0 +\alpha_2)P_T |H_N|^2}{L_{ar}L_{rw}} ~~~~~~\nonumber \\ + \sigma_{w}^{2} > \eta
\bigg)  + \Pr \left(\frac{P_T |h_{aw}|^2}{L_{aw}} + \frac{P_T |H_N|^2}{L_{ar}L_{rw}} + \sigma_{w}^{2} < \eta
\right)\nonumber \\
= \left\{ {\begin{array}{*{20}{c}}
            {1}, &{ \eta< \sigma_{w}^{2} + \delta_1 |h_{aw}|^2},\\
            v_1 , &{\sigma_{w}^{2} + \delta_1 |h_{aw}|^2 \le \eta \le \sigma_{w}^{2} + \delta_2 |h_{aw}|^2 },\\  v_2, &{ \eta > \sigma_{w}^{2} + \delta_2 |h_{aw}|^2},
            \end{array}} \right. ~~~~~~~~ \label{w_n}
\end{align}
where $\delta_1 = \tfrac{(\alpha_0 + \alpha_2)P_T }{L_{aw}}$, $\delta_2 = \tfrac{P_T }{L_{aw}}$,  $v_1 = e^{\frac{\delta_1 |h_{aw}|^2 + \sigma_{w}^{2} - \eta}{\delta_1 \lambda_{N} \varphi^{-1}}}$, $v_2 = 1 + v_1 - e^{\tfrac{\delta_2 |h_{aw}|^2 + \sigma_{w}^{2} - \eta}{\delta_2 \lambda_{N} \varphi^{-1}}}$, and $\varphi = \tfrac{L_{ar} L_{rw}}{L_{aw}}$.
From the perspective of Willie, the minimum DEP can be achieved by choosing an optimal detection threshold $\eta^*$. The minimum DEP and the corresponding optimal detection threshold are provided in the following theorem.

\begin{theo}{The optimal detection threshold  $\eta^*$ and the minimum DEP are respectively given by
\begin{small}
\begin{align}
\eta^* = ~~~~~~~~~~~~~~~~~~~~~~~~~~~~~~~~~~~~~~~~~~~~~~~~~~~~~~~~~~~~~~~~~~~~~~~~~~ \nonumber \\ 
\left\{\!\!\! {\begin{array}{*{20}{c}}
            {\sigma_{w}^{2} + \delta_2|h_{aw}|^2 }, \!\!\!&{ {\rm{if}} |h_{aw}|^2 \ge \tfrac{(\alpha_0 + \alpha_2) \lambda_N}{\alpha_1 \varphi} \ln\left(\tfrac{1}{\alpha_0 + \alpha_2}\right)} \\
            \sigma_{w}^{2} + \tfrac{\delta_1 \lambda_N}{\alpha_1 \varphi}\ln \left(\tfrac{1}{\alpha_ 0+ \alpha_2}\right) ,\!\!\! &{\rm{otherwise} }
            \end{array}} \right.   \label{n_1}
\end{align}
\end{small}
and
\begin{flalign}
\xi_{w,\eta}^{*} =
\left\{ {\begin{array}{*{20}{c}}
            {v_{1}^{*} }, &{ {\rm{if}}~~ |h_{aw}|^2 \ge \frac{(\alpha_0 + \alpha_2) \lambda_N}{\alpha_1 \varphi} \ln\left(\frac{1}{\alpha_0 + \alpha_2}\right)}\\
            v_{2}^{*} , &{\rm{otherwise} }
            \end{array}} \right. , \label{w_n1}
\end{flalign}
where $v_{1}^{*} = e^{-\frac{\alpha_1 \varphi |h_{aw}|^2}{(\alpha_0 + \alpha_2) \lambda_N} }$ and $v_{2}^{*} = 1 - \alpha_1 (\alpha_0 + \alpha_2)^{\frac{1-\alpha_1}{\alpha_1}} \times e^{\frac{\varphi |h_{aw}|^2}{\lambda_N}}$.}
\end{theo}

\begin{IEEEproof}
See Appendix A.
\end{IEEEproof}

{\it{\textbf{Remark~2:}}} The results in Theorem 1 show that both $\eta^*$ and $\xi_{w,\eta}^{*}$ depend on $|h_{aw}|^2$. Taking into account the unavailability of instantaneous CSI of $|h_{aw}|^2$ for the legitimate system, the MADEP of Willie is adopted as the performance metric of the covertness. By using the total probability theorem, the MADEP of Willie can be derived as:
\begin{small}
\begin{align}
\bar{\xi}_{w,\eta}^{*} = \int\nolimits_{\tfrac{(\alpha_0 + \alpha_2)\lambda_N}{\alpha_1 \varphi}\ln \left(\tfrac{1}{\alpha_0 + \alpha_2}\right)}^{+\infty} e^{-\tfrac{\alpha_1 \varphi x}{(\alpha_0 + \alpha_2)\lambda_N}} f_{|h_{aw}|^2} (x) dx~~~~~~~ \nonumber \\ + \int\nolimits_{0}^{\tfrac{(\alpha_0 + \alpha_2)\lambda_N}{\alpha_1 \varphi}\ln \left(\tfrac{1}{\alpha_0 + \alpha_2}\right)} \left[1 - \alpha_1 (\alpha_0 + \alpha_2)^{\tfrac{1-\alpha_1}{\alpha_1}}e^{\tfrac{\varphi x}{\lambda_N}} \right] \nonumber \\ \times  f_{|h_{aw}|^2} (x) dx \nonumber \\ = \tfrac{(\alpha_0 + \alpha_2)\lambda_N }{\alpha_1 \varphi \lambda_{aw}  + (\alpha_0 + \alpha_2)\lambda_N} \left(\alpha_0 + \alpha_2\right)^{1 + \tfrac{(\alpha_0 + \alpha_2)\lambda_N}{\alpha_1 \varphi \lambda_{aw}}} + 1~~~~~~~~~~ \nonumber \\ -  (\alpha_0 + \alpha_2)^{\tfrac{(\alpha_0 + \alpha_2)\lambda_N}{ \alpha_1 \varphi\lambda_{aw}}} -\tfrac{\alpha_1 \lambda_N (\alpha_0 + \alpha_2)^{-1+\tfrac{1}{\alpha_1}}}{\lambda_{aw} \varphi  - \lambda_N}~~~~~~~~~~~~ \nonumber \\ \times \left[(\alpha_0 + \alpha_2)^{\tfrac{(\lambda_N - \varphi \lambda_{aw})(\alpha_0 + \alpha_2)}{\alpha_1 \varphi\lambda_{aw}}} - 1 \right].~~~~~~~~~~~~~~~~~~~~\label{MADEP}
\end{align}
\end{small}
{\it{\textbf{Remark~3:}}}  Considering the case of $\alpha_2 = 0$, in which Alice does not transmit the private signal of Grace. In this case, Alice's transmit power is allocated to the common signal and Bob's private signal. As $\alpha_1 \to 1$, or equivalently $\alpha_0 \to 0$, we have $\alpha_1 P_T \to P_T$, i.e., all the transmit power is allocated to Bob's private signal, which results in $\bar{\xi}_{w, \eta}^{*} \to 0$. On the other hand, as $\alpha_0 \to 1$, or equivalently $\alpha_1 \to 0$, we have $\alpha_0 P_T \to P_T$, i.e., all the transmit power is allocated to public signal, which results in $\bar{\xi}_{w, \eta}^{*} \to 1$. However, Bob and Grace can only obtain the decreased achievable rates from the transmission of public signal, as it will be verified in the simulation results.

{\it{\textbf{Remark~4:}}}  Considering the case of  $\alpha_2 = 1$, i.e., Alice allocates all of its transmit power to transmit the private signal to Grace. For this case, it is easy to see from \eqref{MADEP} that $\bar{\xi}_{w, \eta}^{*} = 1$, such that Willie cannot make a correct judgment based on Alice's common and private transmissions to Grace. Although the full covertness ($\bar{\xi}_{w, \eta}^{*} = 1$) is guaranteed in this case, the achievable rate of Bob is zero due to the fact  $\alpha_0 = \alpha_1 = 0$.

\section{Covert Rate Maximization}

\subsection{Problem Formulation and Parameters Optimization}

In order to maximize the covert rate of Bob subjecting to the  covertness constraint and Grace's QoS requirements, the transmit power allocation, rate allocation, and STAR-RIS reflection/transmission beamforming need to be jointly optimized. Specifically, to guarantee the required covertness, the MADEP of Willie cannot be less than a predefined level $1-\varepsilon$, where $\varepsilon$ is a positive constant representing a certain level of the covertness. To guarantee Grace's QoS requirements, the achievable rate of Grace needs to be no less than the pre-defined target rate $R_{g}^{\rm{min}}$, i.e., $R_g \ge R_{g}^{\rm{min}}$. Therefore, the optimization problem that maximizes Bob's covert rate can be formulated as:
\begin{subequations}
\begin{align}
({\rm{P1}}):~~ & \!\!\!\! \mathop{{\rm{max}}}  \limits_{\alpha_0, \alpha_1, \alpha_2, \beta_1, \beta_2, {\bf{\Theta}}_r, {\bf{\Theta}}_{t}}  R_b & \label{P1a}\\
{\rm{s.t.}}~~ &  \alpha_0 + \alpha_1 + \alpha_2 \le 1, ~0\le \alpha_1, \alpha_2, \alpha_3 \le 1, &  \label{P1_alpha} \\
& \beta_1 + \beta_2 \le 1, ~0 \le \beta_1, \beta_2 \le 1, &  \label{P1_beta} \\
& R_{b}^{c}\ge 0, ~R_{g}^{c} \ge 0,  &\label{P1_Rc}
\\
& R_g \ge R_{g}^{{\rm{min}}},  &  \label{P1_Grace_rate}\\
& \bar{\xi}_{w,\eta}^{*} \ge 1 - \varepsilon , ~\varepsilon \in [0,1], & \label{P1_covertness} \\
& 0 \le \theta_{n}^{r} \le 2\pi, ~n = 1,\cdots, K_n,   & \label{P1_Theta_r} \\
& 0 \le \theta_{m}^{t} \le 2\pi, ~m = 1,\cdots, K_m.  & \label{P1_Theta_t}
\end{align}
\end{subequations}
In problem (P1),  \eqref{P1_alpha} and \eqref{P1_beta} are the constraints on the transmit  power allocation and rate allocation, respectively, $\eqref{P1_Rc}$ denotes the constraints on the  common rate for Bob and Grace, $\eqref{P1_Grace_rate}$ denotes Grace's QoS  requirement, and \eqref{P1_covertness} guarantees a large value of MADEP to realize the covertness. Moreover, \eqref{P1_Theta_r} and \eqref{P1_Theta_t} are the constraints on the STAR-RIS reflection and transmission beamforming, respectively.

Since the system parameters in the objective function of (P1) and constraints \eqref{P1_Rc}, \eqref{P1_Grace_rate}, and \eqref{P1_covertness} are highly coupled with each other, it is impossible to obtain the optimal system parameters directly. In addition, the expression ${\bar \xi_{w, \eta}^{*}}$ in the constraint \eqref{P1_covertness} is too complex to gain the insights on the covert communication design. To tackle the above challenges, we first decouple the original problem (P1) into three sub-problems, namely, power allocation optimization problem, rate allocation optimization problem, and STAR-RIS reflection and transmission beamforming optimization problem, and propose an efficient AO algorithm to determine the optimized system parameters.

For any given variables $\beta_1$, $\beta_2$, ${\bf{\Theta}}_r$, and ${\bf{\Theta}}_t$, it can be observed that the value of the objective function in problem (P1) depends only on the power allocation factors $\alpha_0$, $\alpha_1$, and $\alpha_2$. Therefore, problem (P1) can be reformulated as: 
\begin{subequations}
\begin{align}
({\rm{P2}}):~& \mathop{{\rm{max}}}\limits_{\alpha_0, \alpha_1, \alpha_2} R_b &  \label{P2a} \\
~~~~{\rm{s.t.}}~ &\alpha_0 + \alpha_1 + \alpha_2 \le 1,~ 0 \le \alpha_0, \alpha_1, \alpha_2 \le 1,&  \label{P2b}
\\
&\beta_2 \log_2 \left(1 + \tfrac{\alpha_0 P_{T} |Z_{ag}|^2}{(\alpha_1+\alpha_2)P_{T}|Z_{ag}|^2 + \sigma_{g}^{2}} \right)& \nonumber \\&  + \log_2\left(1+\tfrac{\alpha_2 P_{T}|Z_{ag}|^2}{\alpha_1 P_{T} |Z_{ag}|^2 + \sigma_{g}^{2}} \right) \ge R_{g}^{\rm{min}},&
 \label{P2e} \\
&\bar{\xi}_{w,\eta}^{*} \ge 1 - \varepsilon , ~\varepsilon \in [0,1]. &   \label{P2f}
\end{align}
\end{subequations}
In problem (P2), we find that the variables $\alpha_0$, $\alpha_1$, and $\alpha_2$ are coupled with each other in the objective function and constraints, while constraint \eqref{P2f} is very complex. Thus,  it is challenging to obtain the optimal solution of problem (P2) directly. To deal with these challenges, we design an AO algorithm to obtain the optimal $\alpha_0$, $\alpha_1$, and $\alpha_2$. The reason for the designed AO algorithm is that we can obtain the optimal $\alpha_{1}^{*}$ and $\alpha_{2}^{*}$ in the closed-forms when $\alpha_0$ is given and vise verse. In addition, we have the following proposition on the optimal transmit power allocation. 

\textbf{Proposition 1}: The optimal solution to problem (P2) can be achieved only when $\alpha_0 + \alpha_1 + \alpha_2 = 1$.

\begin{IEEEproof}
The proof is given in Appendix B.
\end{IEEEproof}

{\it{\textbf{Remark~5:}}} According to the results in Proposition 1, we have  $\alpha_0  + \alpha_2 = 1 - \alpha_1$ if the maximum $R_b$ is achieved. By substituting $\alpha_0  + \alpha_2 = 1 - \alpha_1$ into \eqref{MADEP}, the MADEP of Willie can be rewritten as:
\begin{flalign}
\bar{\xi}_{w,\eta}^{*} = \frac{(1 - \alpha_1)\lambda_N }{\alpha_1 \varphi \lambda_{aw}  + (1 - \alpha_1)\lambda_N} \left(1 - \alpha_1\right)^{1 + \frac{(1 - \alpha_1)\lambda_N}{\alpha_1 \varphi \lambda_{aw}}}~~~~~~~~ \nonumber \\ + 1 -  (1 - \alpha_1)^{\frac{(1 - \alpha_1)\lambda_N}{ \alpha_1 \varphi\lambda_{aw}}} ~~~~~~~~~~~~~~~~~~~~~~~~~~~~~~~~~~ \nonumber \\ -\tfrac{\alpha_1 \lambda_N (1 - \alpha_1)^{-1+\tfrac{1}{\alpha_1}}}{\lambda_{aw} \varphi  - \lambda_N} \left[(1 - \alpha_1)^{\frac{(\lambda_N - \varphi \lambda_{aw})(1 - \alpha_1)}{\alpha_1 \varphi\lambda_{aw}}} - 1 \right].\label{MADEP_2}
\end{flalign}
According to \eqref{MADEP_2}, from the perspective of the transmit power allocation, $\bar{\xi}_{w,\eta}^{*}$ is only determined by $\alpha_1$ when the maximum $R_b$ is achieved, whereas the relative ratio between $\alpha_0$ and $\alpha_2$ does not affect $\bar{\xi}_{w,\eta}^{*}$ in such a case.

\begin{algorithm}[t]
\caption{AO Algorithm of Optimizing Transmit Power Allocation}
 \label{alg11}
 \begin{algorithmic}[1]
  \STATE Initialize a feasible $\alpha_0$.
  \STATE $\iota \gets 0$
  \REPEAT
  \STATE Given $\alpha_{0}^{\iota }$, update $\alpha_{1}^{\iota+1}$ and $\alpha_{2}^{\iota+1}$ according to \eqref{a_1} and \eqref{a_2}.
  \STATE Given $\alpha_{1}^{\iota+1 }$ and $\alpha_{2}^{\iota+1}$, update $\alpha_{0}^{\iota+1}$ according to $\alpha_{0}^{\iota+1} = 1-\alpha_{1}^{\iota+1}-\alpha_{2}^{\iota+1}$.
  \UNTIL $R_{b}^{\iota+1} - R_{b}^{\iota} < \zeta_1$.
 \end{algorithmic}
\end{algorithm}

For any given $\alpha_0$, we have $\alpha_1+\alpha_2 = 1-\alpha_0$ according  to the results of Proposition 1. Consequently, in constraint \eqref{P2e}, Grace's common rate $R_{g}^{c}$ can be determined once $a_0$ is given. Then, constraint \eqref{P2e} can be simplified to $R_{g}^{c} + \log_2 \left(1 + \tfrac{(1-\alpha_0-\alpha_1)P_T |Z_{ag}|^2}{\alpha_1 P_T |Z_{ag}|^2 + \sigma_{g}^{2}} \right) \ge R_{g}^{\rm{min}}$. For constraint \eqref{P2e}, we have the following observations: 1) In the case of  $R_{g}^{c} \ge R_{g}^{\rm{min}}$, Grace's QoS requirements have been satisfied because its common rate is no less than the target rate. Thus, Alice does not need to allocate transmit power to Grace's private stream. 2) In the case of  $R_{g}^{c} < R_{g}^{\rm{min}}$, Alice needs to allocate more transmit power to Grace's private streams to satisfy Grace's QoS requirements. Moreover, for any given $\alpha_0$, $R_b$ is a monotonically increasing function with respect to $\alpha_1$. According to the above observations, for any given $\alpha_0$, the optimal $\alpha_{1}^{*}$ and $\alpha_{2}^{*}$  can be respectively expressed as
\begin{flalign}
\alpha_{1}^{*} =
\left\{ {\begin{array}{*{20}{c}}
            {  {\rm{min}} \left\{1-\alpha_0,~ \Xi_1 \right\}
            }, &{ {\rm{if}}~~ R_{g}^{c}} \ge R_{g}^{\rm{min}} \\
            { {\rm{min}} \left\{1-\alpha_0,~ \Xi_1,~ \Xi_2 \right\}
            }, &{\rm{otherwise} }
            \end{array}} \right. \label{a_1}
\end{flalign}
and
\begin{flalign}
\alpha_{2}^{*} =
\left\{ {\begin{array}{*{20}{c}}
            {  0
            },~ &{ {\rm{if}}~~ R_{g}^{c}} \ge R_{g}^{\rm{min}} \\
            { \Xi_3
            }, &{\rm{otherwise} }
            \end{array}} \right. ,\label{a_2}
\end{flalign}
where $\Xi_2 = \tfrac{(1-\alpha_0) P_T |Z_{ag}|^2 - \big(2^{R_{g}^{\rm{min}} - R_{g}^{c}}-1 \big) \sigma_{g}^{2}}{P_T |Z_{ag}|^{2} 2^{R_{g}^{\rm{min}}-R_{g}^{c}}}$, $\Xi_3 = \tfrac{\left(2^{R_{g}^{\rm{min}}-R_{g}^{c}}-1 \right) \left(\alpha_1 P_T |Z_{ag}|^2 +\sigma_{g}^{2} \right)}{P_T|Z_{ag}|^2}$, and $\Xi_1 = \bar{\xi}_{w,\eta}^{*(-1)} (1-\varepsilon)$ is the inverse function of $\bar{\xi}_{w,\eta}^{*}$, which can be computed numerically using a mathematical tool, such as MATLAB.
Now, the optimal value of $\alpha_0^*$ can be obtained as $\alpha_{0}^{*} = 1- \alpha_{1}^{*} - \alpha_{2}^{*}$ for any given $\alpha_1$ and $\alpha_2$. Based on the results of Proposition 1, Algorithm 1 containing the AO approach is proposed to solve the problem (P2).

For any given $\alpha_{0}$, $\alpha_{1}$, $\alpha_{2}$, ${\bf{\Theta}}_r$,  and ${\bf{\Theta}}_t$, the objective function in problem (P1) and the corresponding constraints are determined by the variables $\beta_1$ and $\beta_2$ only. Thus, problem (P1) can be equivalently transformed to
\begin{subequations}
\begin{align}
({\rm{P3}}):~&\mathop{{\rm{max}}}\limits_{\beta_1, \beta_2}~R_b &\label{P4a} \\
~~~~{\rm{s.t.}}~ &\beta_2 \log_2 \left( 1 + \tfrac{\alpha_0 P_T |Z_{ag}|^2}{(\alpha_1 + \alpha_2)P_T|Z_{ag}|^2 + \sigma_{g}^{2}} \right)& \nonumber \\& + \log_2 \left( 1 + \tfrac{\alpha_2 P_T |Z_{ag}|^2}{\alpha_1 P_T |Z_{ag}|^2 + \sigma_{g}^{2}} \right) \ge R_{g}^{\rm{min}}.&  \label{P4b}
\end{align}
\end{subequations}
In problem (P3), constraint \eqref{P4b} can be further expressed as $\beta_2 \ge \Xi_4$, where $\Xi_4 = \tfrac{R_{g}^{\rm{min}} - \log_2 \left( 1 + \tfrac{\alpha_2 P_T |Z_{ag}|^2}{\alpha_1 P_T |Z_{ag}|^2 + \sigma_{g}^{2}} \right) }{\log_2 \left( 1 + \tfrac{\alpha_0 P_T |Z_{ab}|^2}{(\alpha_1 + \alpha_2)P_T|Z_{ab}|^2 + \sigma_{b}^{2}} \right)}$. Considering that $R_b$ is a monotonically decreasing function with respect to $\beta_2$, the optimal solution to problem (P3) is given by $\beta_{1}^{*} = 1-\beta_{2}^{*}$ and $\beta_{2}^{*} =\Xi_4$.

For any given $\alpha_0$, $\alpha_1$, $\alpha_2$, $\beta_1$, and $\beta_2$, problem (P1) turns to be the covert rate maximization problem with respect to ${\bf{\Theta}}_r$ and ${\bf{\Theta}}_t$. To make the corresponding optimization feasible, we introduce a series of variables and conduct the problem formulation accordingly. Specifically, let ${\bf{u}}_r = [u_{1}^{r},\cdots,u_{n}^{r}]^{H}$ with $u_{n}^{r} = e^{j\theta_{n}^{r}}$, ${\bf{u}}_t = [u_{1}^{t},\cdots,u_{m}^{t}]^{H}$ with $u_{m}^{t} = e^{j\theta_{m}^{t}}$, where $|u_{n}^{r}|^2 = 1 $, $\forall_n$ and $|u_{m}^{t}|^2 = 1$, $\forall_m$. Then, we construct $ \bar{{\bf{u}}}_r = \left[ {\bf{u}}_{r}^{{\rm{T}}},~1  \right]^{{\rm{T}}}  $, $\textbf{H}_g = \Lambda_g \Lambda_{g}^{H}$, and
\begin{eqnarray}
\textbf{H}_b = \left[\begin{array}{cc}
    \Lambda_b \Lambda_{b}^{H} & \Lambda_b \nu_b \\
    \nu_b \Lambda_{b}^{H} & 0
\end{array} \right],
\end{eqnarray}
where $\Lambda_g = \frac{{\rm diag}({\bf{h}}_{ar_2}^{H}){\bf{h}}_{rg}}{\sqrt{L_{ar} L_{rg}}}$, $\Lambda_b = \frac{{\rm diag}({\bf{h}}_{ar_1}^{H}){\bf{h}}_{rb}}{\sqrt{L_{ar} L_{rb}}}$ and $\nu_b = \frac{h_{ab}}{\sqrt{L_{ab}}}$.
Now, we can rewrite the composite channel gains as  $|Z_{ab}|^2 = |{\bf{u}}_{r}^{H} \Lambda_b + \nu_b|^2 = \bar{{\bf{u}}}_{r}^{H} {\bf{H}}_b \bar{{\bf{u}}}_r + |\nu_b|^2 = {\rm{Tr}} ({\bf{H}}_b {\bf{U}}_r) + |\nu_b|^2$ and $|Z_{ag}|^2 = |{\bf{u}}_{t}^{H} \Lambda_g|^2 = {\bf{u}}_{t}^{H} {\bf{H}}_g {\bf{u}}_t = {\rm{Tr}} ({\bf{H}}_g {\bf{U}}_t)$, where ${\bf{U}}_r = \bar{{\bf{u}}}_r \bar{{\bf{u}}}_{r}^{H} \succeq {\bf 0}$ and  ${\bf{U}}_t = {{\bf{u}}}_t {{\bf{u}}}_{t}^{H} \succeq {\bf 0}$ are two semi-definite matrices satisfying ${\rm{rank}}({\bf{U}}_r) = {\rm{rank}}({\bf{U}}_t) = 1$.
Since maximizing $R_b=\beta_1 R_{g, s_0(q)} + R_{b, s_1(q)}$ is equivalent to maximize $\beta_1 R_{g, s_0(q)}$ and $R_{b, s_1(q)}$, respectively, while maximizing $R_{g, s_0(q)}$ and $R_{b, s_1(q)}$ are equivalent to maximize the channel gains $|Z_{ag}|^2$ and $|Z_{ab}|^2$, respectively, the maximum $R_b$ can be obtained by maximizing $|Z_{ab}|^2 + |Z_{ag}|^2$ with its equivalent form as ${\rm{Tr}}({\bf{H}}_b {\bf{U}}_r) + {\rm{Tr}}({\bf{H}}_g {\bf{U}}_t)$. 
Thus, for any given $\alpha_0$, $\alpha_1$, $\alpha_2$, $\beta_1$, and $\beta_2$, the covert rate maximization problem in the terms of optimizing the STAR-RIS reflection and transmission beamforming can be equivalently formulated as:
\begin{subequations}
\begin{align}
({\rm{P4}}):~&\mathop{{\rm{max}}}\limits_{{\bf{U}}_r, {\bf{U}}_t} {\rm{Tr}}({\bf{H}}_b {\bf{U}}_r) + {\rm{Tr}}({\bf{H}}_g {\bf{U}}_t) &\label{P7a} \\
{\rm{s.t.}}~ &{\rm{Tr}}({\bf{H}}_b {\bf{U}}_r)+|v_b|^2 > {\rm{Tr}}({\bf{H}}_g {\bf{U}}_t),& \label{P7b}  \\ &{\rm{Tr}}({\bf{H}}_g {\bf{U}}_t)\ge R_{g}^{\rm{min}(-1)}, &\label{P7c}
  \\
&{\bf{U}}_{r,(n,n)} = 1,~ n = 1,\cdots,K_n,& \label{P7e}
 \\
&{\bf{U}}_{t,(m,m)} = 1,~ m = 1,\cdots,K_m,&\label{P7f} \\
&{\bf{U}}_r \succeq {\bf 0}, ~{\bf{U}}_t \succeq {\bf 0}, &\label{P7g} \\
&{\rm{rank}}({\bf{U}}_r) = {\rm{rank}}({\bf{U}}_t) = 1.&\label{P7h}
\end{align}
\end{subequations}
In problem (P4), constraint \eqref{P7c} represents Grace's QoS requirements with $ R_{g}^{\rm{min}(-1)}$ denoting the inverse function of  $ R_{g}^{\rm{min}} (|Z_{ag}|^2)$.
However, problem (P4) is still non-convex due to the rank-one constraint \eqref{P7h}. For tackle this kind of non-convex problems, the general-rank solution can be obtained by using the semidefinite relaxation method followed by the Gaussian randomization to construct the rank-one solution approximately \cite{9524501,8647620}. Unfortunately, this kind of methods usually leads to large errors. To solve problem (P4), we adopt a PSCA approach to obtain an optimal solution to problem (P4), specifically, we first transform the rank-one constraint to a penalty term to be included in the objective function \cite{9668964} and then invoke successive convex approximation (SCA) \cite{7547360} to obtain the optimal solution. To this end, we introduce an equivalent inequality constraint as:
\begin{eqnarray}
\|{\bf{U}}_k\|_* - \|{\bf{U}}_k\|_2  \ge 0,~ k \in \{r, t \} , \label{Uys}
\end{eqnarray}
where $\|{\bf{U}}_k\|_*$ is the nuclear norm defined as the sum of singular values of ${\bf{U}}_k$ and $\|{\bf{U}}_k\|_2$ is the spectral norm defined as the largest singular value of ${\bf{U}}_k$. When ${\rm{rank}}({\bf{U}}_r) = {\rm{rank}}({\bf{U}}_g) = 1$, we have $\|{\bf{U}}_k\|_* - \|{\bf{U}}_k\|_2  = 0$. Otherwise, $\|{\bf{U}}_k\|_* - \|{\bf{U}}_k\|_2  > 0$. In order to obtain a rank-one matrix solution, a penalty term is added into the objective function considering the equivalence between $\|{\bf{U}}_k\|_* - \|{\bf{U}}_k\|_2  = 0$ and ${\rm{rank}}({\bf{U}}_r) = {\rm{rank}}({\bf{U}}_t) = 1$. Then, problem (P4) is transformed as:
\begin{subequations}
\begin{align}
({\rm{P5}}):~& \mathop{{\rm{max}}}\limits_{{\bf{U}}_r, {\bf{U}}_t} {\rm{Tr}}({\bf{H}}_b {\bf{U}}_r) + {\rm{Tr}}({\bf{H}}_g {\bf{U}}_t) & \nonumber \\ &~~~~~~~~~~ - \frac{1}{\epsilon} \sum_{k\in \{r, t\}} \left(\|{\bf{U}}_k\|_* - \|{\bf{U}}_k\|_2 \right) & \label{P8a} \\
~ &~~~~{\rm{s.t.}}~~ \eqref{P7b},~ \eqref{P7c},~ \eqref{P7e},~  \eqref{P7f},~  \eqref{P7g}. &   \label{P8b}
\end{align}
\end{subequations}
where $\epsilon$ is a control factor. Nevertheless, the objective function in problem (P5) is non-convex due to the introduction of the penalty term. To solve this problem, we use the first-order Taylor expansion at point ${\bf{U}}_{k}^{n}$ and invoke the SCA to obtain an upper bound on $- \|{\bf{U}}_k\|_2$ as:
\begin{flalign}
&- \|{\bf{U}}_k\|_2  \le \widehat{{\bf{U}}}_{k}^{n} \triangleq & \nonumber \\ &~~~~~~~~~~~~~~  -\|{\bf{U}}_{k}^{n}\|_2 - {\rm{Tr}} \left( {\bf{v}}_{\max,k}^{n} ({\bf{v}}_{\max,k}^{n})^H ({\bf{U}}_{k} -{\bf{U}}_{k}^{n}) \right) ,& \label{Usj}
\end{flalign}
where $n$ denotes the iteration index and ${\bf{v}}_{\max,k}^{n}$ is the eigenvector corresponding to the largest eigenvalue of ${\bf{U}}_{k}^{n}$. Then, problem (P5) can be approximated as:
\begin{subequations}
\begin{align}
({\rm{P6}}):~&\mathop{{\rm{max}}}\limits_{{\bf{U}}_r, {\bf{U}}_t} {\rm{Tr}}({\bf{H}}_b {\bf{U}}_r) + {\rm{Tr}}({\bf{H}}_g {\bf{U}}_t) & \nonumber \\ &~~~~~~~~~~ - \frac{1}{\epsilon} \sum_{k\in \{r, t\}} \left(\|{\bf{U}}_k\|_* + \widehat{{\bf{U}}}_{k}^{n} \right)&    \label{P90a} \\
~ &~~~{\rm{s.t.}}~~\eqref{P7b},~ \eqref{P7c},~ \eqref{P7e},~  \eqref{P7f},~  \eqref{P7g}. &   \label{P90b}
\end{align}
\end{subequations}

\begin{algorithm}[t]
 %\textsl{}\setstretch{1.8}
% \renewcommand{\algorithmicrequire}{\textbf{Input:}}
% \renewcommand{\algorithmicensure}{\textbf{Output:}}
 \caption{AO Algorithm of Maximizing STAR-RIS-RSMA Covert Rate}
 \label{alg1}
 \begin{algorithmic}[1]
%     \REQUIRE System bound $B$, user information $P_a$, $\hat{R}_a$, $N_a$, $N_b$, $\sigma^2$, channel information $\ell_a$, $h_a$, $\ell_b$, $h_b$.
%  \ENSURE Power allocation $P_{b,1}$, $P_{b,2}$, rate splitting factor $\beta$, power allocation factor $\alpha$, and time phase $t_2$.
  \STATE Initialize the feasible $\beta_{1}^{0}, \beta_{2}^{0}, {\bf{\Theta}}_{r}^{0}$, and ${\bf{\Theta}}_{t}^{0}$.
  \STATE ${\bf{\Theta}}_{r}^{0}$, and ${\bf{\Theta}}_{t}^{0}$ can be equivalently transformed to ${\bf{U}}_{r}^{0}$ and ${\bf{U}}_{t}^{0}$.
  \STATE $\ell \gets 0$
  \REPEAT
  \STATE Given $\beta_{1}^{\ell}, \beta_{2}^{\ell}, {\bf{U}}_{r}^{\ell}$ and ${\bf{U}}_{t}^{\ell}$, solve the problem (P2) by \textbf{Algorithm 1} to obtain $\alpha_{0}^{\ell+1}, \alpha_{1}^{\ell+1}$, and $\alpha_{2}^{\ell+1}$.
  \STATE Given $\alpha_{0}^{\ell+1}, \alpha_{1}^{\ell+1}$, $\alpha_{2}^{\ell+1}$, ${\bf{U}}_{r}^{\ell}$ and ${\bf{U}}_{t}^{\ell}$, solve the problem (P3) to obtain $\beta_{1}^{\ell+1}$ and $\beta_{2}^{\ell+1}$.
  \STATE $n \gets 0$
  \REPEAT
  \STATE Given $\alpha_{0}^{\ell+1}, \alpha_{1}^{\ell+1}$, $\alpha_{2}^{\ell+1}$,  $\beta_{1}^{\ell+1}$, and $\beta_{2}^{\ell+1}$ solve the problem (P6) to obtain ${\bf{U}}_{r}^{n+1}$ and ${\bf{U}}_{t}^{n+1}$.
  \STATE Update $n = n + 1$.
  \STATE ${\bf{U}}_{r}^{0} \gets {\bf{U}}_{r}^{n}$, ${\bf{U}}_{t}^{0} \gets {\bf{U}}_{t}^{n}$.
  \STATE $\epsilon \gets c_1 \epsilon$.
  \UNTIL $\sum_{k\in \{r, t\}} \left(\|{\bf{U}}_k\|_{*}^{n} - \|{\bf{U}}_k\|_{2}^{n} \right) \le \zeta_2$.
  \STATE ${\bf{U}}_{r}^{0} \gets {\bf{U}}_{r}^{\ell+1}$, ${\bf{U}}_{t}^{0} \gets {\bf{U}}_{t}^{\ell+1}$.
  \STATE Update $\ell = \ell + 1$.
  \UNTIL $R_{b}^{\ell+1} - R_{b}^{\ell} < \zeta_3$.
 \end{algorithmic}
\end{algorithm}
Now, problem (P6) is a semidefinite programming (SDP) problem, which can be solved by using a convex optimization tool such as CVX \cite{9524501,8647620}.
In addition, we have the following insights on the choice of the parameter $\epsilon$: If the parameter $\epsilon \to 0$ ($\frac{1}{\epsilon} \to \infty$), the rank of the matrix ${\bf{U}}_r$ and ${\bf{U}}_t$ will be definitely 1, whereas the objective function is dominated by the penalty term. Thus, the optimal solution to maximize the covert rate can hardly be obtained. To address this issue, we first define a larger initial value for $\epsilon$ to obtain a good starting point for maximizing the covert rate. Then, the value of $\epsilon$ is iteratively decreased by setting $\epsilon = c_1 \epsilon,~ 0 \le c_1 \le 1$ until the penalty term satisfies a predefined threshold $\zeta_2$. The whole procedure of the AO approach is given in Algorithm 2. The convergence of Algorithm 2 is characterized by the following proposition.

\textbf{Proposition 2}: The developed Algorithm 2 is guaranteed to converge.

\begin{IEEEproof}
The proof is given in Appendix C.
\end{IEEEproof}

In Algorithm 1, the optimal solution is solved in each iteration by a closed-form expression, thus the complexity of Algorithm 1 is $\mathcal{O} \left(I_0 (3)\right)$, where $I_0$ denotes the number of iterations of the AO loop. The complexity of Algorithm 1 is almost negligible compared to solving the SDP problem in Algorithm 2. The complexity of Algorithm 2 mainly comes from solving the SDP problem (P6) during the iterations. Moreover, the objective function of (P6) includes a penalty term, which adds the additional complexity to Algorithm 2. The complexity of solving (P6) is characterized by  $\mathcal{O} \left(I_2 \left((K_n+1)^{3.5} +K_{m}^{3.5} \right)\right)$ \cite{8723525}, where $I_2$ is the number of iterations with respect to the penalty term. Then, the overall complexity of Algorithm 2 can be characterized by $\mathcal{O} \left(I_1 I_2 \left((K_n+1)^{3.5} +K_{m}^{3.5} \right)\right)$, where $I_1$ represents the number of iterations of the AO out-loop.

\subsection{Benchmark Scheme}

This section discusses the covert performance of the STAR-RIS-NOMA system. In the considered STAR-RIS-NOMA system, the power-domain NOMA is used to realize the transmissions from Alice to Bob and Grace \cite{9524501,9146329}. Specifically, the messages $\tilde s_b(q)$ and $\tilde s_g(q)$ are encoded to $s_b(q)$ and $s_g(q)$, respectively. The SIC decoding order at Bob is $ s_g(q) \to s_b(q)$, meaning that $s_g(q)$ is decoded before $s_b(q)$. At Grace, $s_g(q)$ is decoded treating the signal related to $s_b(q)$ as noise. To ensure successful SIC at the receiver and realize Bob's covert communications, more transmit power is allocated to transmit $ s_g(q)$ at Alice \cite{9079918,9524501,9146329}. Taking into account the transmit power allocation factors $\bar{\alpha}_1$ and $\bar{\alpha}_2$ for $s_b(q)$ and $s_g(q)$ respectively. These factors satisfy $\bar{\alpha}_1 +\bar{\alpha}_2 \le 1$ and $0 \le \bar{\alpha}_1, \bar{\alpha}_2 \le 1$. Then,
the achievable rate of Bob from the transmission of $s_g(q)$ can be written as: 
\begin{eqnarray}
R_{b,s_g (q)}^{^{\rm NOMA}} = \log_2 \left(1 + \frac{\bar{\alpha}_2 P_T |Z_{ab}|^2}{\bar{\alpha}_1 P_T|Z_{ab}|^2 + \sigma_{b}^{2}}\right).
\end{eqnarray} 
Similarly, Bob's covert rate obtained from the transmission of $s_b(g)$ can be expressed as:
\begin{eqnarray}
R_{b,s_b (q)}^{^{\rm NOMA}} = \log_2 \left(1 + \frac{\bar{\alpha}_1 P_T |Z_{ab}|^2}{ \sigma_{b}^{2}}\right).
\end{eqnarray}
As such, the achievable rate of Grace is given by 
\begin{eqnarray}
R_{g,s_g (q)}^{^{\rm NOMA}} = \log_2 \left(1 + \frac{\bar{\alpha}_2 P_T |Z_{ag}|^2}{\bar{\alpha}_1 P_T|Z_{ag}|^2 + \sigma_{g}^{2}}\right).
\end{eqnarray}

Similar to the STAR-RIS-RSMA system, Willie also adopts the Neyman-Pearson criterion to realize the binary decision. It can be shown that the AMDEP achieved by Willie has a similar form as that in the STAR-RIS-RSMA system. Then, the covert rate maximization problem in the STAR-RIS-NOMA system can be formulated as: 
\begin{subequations}
\begin{align}
({\rm{P7}}):~&\mathop{{\rm{max}}}\limits_{\bar{\alpha}_1, \bar{\alpha}_2, {\bf{\Theta}}_r  {\bf{\Theta}}_t}  R_{b,s_b (q)}^{^{\rm NOMA}}  &\label{P9a}\\
{\rm{s.t.}}~ &\bar{\alpha}_1 + \bar{\alpha}_2 \le 1,~  \bar{\alpha}_1 \le \bar{\alpha}_2,&\label{P9b}\\&
R_{b,s_b (q)}^{^{\rm NOMA}} > R_{g,s_g (q)}^{^{\rm NOMA}},& \label{P9c}
\\
&R_{g,s_g(q)}^{^{\rm NOMA}} \ge R_{g}^{{\rm{min}}},  &\label{P9d}\\
&\bar{\xi}_{w,\eta}^{*} \ge 1 - \varepsilon ,~ \varepsilon \in [0,1],& \label{P9e} \\
&0 \le \theta_{n}^{r} \le 2\pi,~ n = 1,\cdots,K_n, &\label{P9f} \\
&0 \le \theta_{m}^{t} \le 2\pi,~ m = 1,\cdots,K_m.&\label{P9g}
\end{align}
\end{subequations}
In problem (P7),  \eqref{P9b} and \eqref{P9c} represent the transmit power allocation and channel gain constraints, respectively,   constraint \eqref{P9c} ensures that a successful SIC is realized at Bob. Moreover, the rest of the constraints are similar to those in problem (P1). We can solve problem (P7) in a similar way as that of solving problem (P1) by using the AO approach.  Specifically, for any given ${\bf{\Theta}}_r$ and ${\bf{\Theta}}_t$, the objective function of problem (P7) is determined by $\{\bar\alpha_1, \bar\alpha_2 \}$ only, which results in the following sub-problem:
\begin{subequations}
\begin{align}
({\rm{P8}}):~&\mathop{{\rm{max}}}\limits_{\bar{\alpha}_1, \bar{\alpha}_2}R_{b,s_b(q)} ^{^{\rm NOMA}}&\label{P10a}\\
{\rm{s.t.}}~ &\bar{\alpha}_1 + \bar{\alpha}_2 \le 1,~  \bar{\alpha}_1 \le \bar{\alpha}_2,&\label{P10b}\\
&\bar{\alpha}_2P_{T}|Z_{ag}|^2 \ge \gamma_{\rm th}\big(\bar{\alpha}_1 P_{T}|Z_{ag}|^2 + \sigma_{g}^{2}\big),& \label{P10c}
\\
&\bar{\xi}_{w,\eta}^{*} \ge 1 - \varepsilon ,~ \varepsilon \in [0,1],&\label{P10d}
\end{align}
\end{subequations}
where $\gamma_{\rm th} = 2^{R_{g}^{\rm{min}}}-1$.

\textbf{Proposition 3}: (P8) can be optimized only when $\bar{\alpha}_1 + \bar{\alpha}_2 = 1$.

\begin{IEEEproof}
The proof is similar to that in Appendix B.
\end{IEEEproof}

Based on the results of Proposition 3, we substitute  ${\bar \alpha_2} = 1-{\bar\alpha_1}$ in the constraints of  problem (P8). Then, problem (P8) can be simplified as:
\begin{subequations}
\begin{align}
({\rm{P9}}):~&\mathop{{\rm{max}}}\limits_{\bar{\alpha}_1, \bar{\alpha}_2}R_{b,s_b(q)}^{^{\rm NOMA}}& \label{P11a} \\
~~~~{\rm{s.t.}}~ &\bar{\alpha}_1 \le {\rm{min}} \left\{ \frac{1}{2}, ~\bar{\xi}_{w,\eta}^{*(-1)} (1-\varepsilon) ,~ \Xi_5 \right\}, &\label{P11b}
\end{align}
\end{subequations}
where $\Xi_5 = \frac{P_{T}|Z_{ag}|^2 - \gamma_{\rm th} \sigma_{g}^{2}}{(1+\gamma_{\rm th})|Z_{ag}|^2 P_{T}}$ is obtained from constraint \eqref{P10c}. Since $R_{g, s_b (q)}^{^{\rm NOMA}}$ is a monotonically increasing function with respect to $\bar{\alpha}_1$, the optimal solution of problem (P9) can be expressed as $\bar{\alpha}_{1}^{*} = {\rm{min}} \big\{ \frac{1}{2},~ \bar{\xi}_{w,\eta}^{*(-1)} (1-\varepsilon) ,~ \Xi_5 \big\}$ and $\bar{\alpha}_{2}^{*} = 1-\bar{\alpha}_{1}^{*}$.

For any given $\bar\alpha_1$ and $\bar\alpha_2$, problem (P8) reduces to only optimize the STAR-RIS reflection and transmission beamforming. Similarly to the problem (P6), a PSCA method is employed to guarantee the rank-one solution and the covert rate maximization problem can be formulated as:
\begin{subequations}
\begin{align}
({\rm{P10}}):~&\mathop{{\rm{max}}}\limits_{\bar{{\bf{U}}}_r, \bar{{\bf{U}}}_t} {\rm{Tr}}(\bar{{\bf{H}}}_b \bar{{\bf{U}}}_r)  - \frac{1}{\epsilon} \sum_{k\in \{r, t\}} \left(\|\bar{{\bf{U}}}_k\|_* + \widehat{\bar{{\bf{U}}}}_{k}^{n} \right)& \label{P12a} \\
{\rm{s.t.}}~ &{\rm{Tr}}(\bar{{\bf{H}}}_b \bar{{\bf{U}}}_r) + |\nu_b|^2 \ge {\rm{Tr}}(\bar{{\bf{H}}}_g \bar{{\bf{U}}}_t), &\label{P12b} \\
&{\rm{Tr}}(\bar{{\bf{H}}}_g \bar{{\bf{U}}}_t) > \frac{ \gamma_{\rm th} \sigma_{g}^{2}}{(\bar{\alpha}_2-\gamma_{\rm th} \bar{\alpha}_1) P_{T}}
,& \label{P12c}
  \\
& \; \bar{{\bf{U}}}_{r,(n,n)} = 1,~ n = 1,\cdots,K_n,& \label{P12d}
 \\
& \;  \bar{{\bf{U}}}_{t,(m,m)} = 1,~ m = 1,\cdots,K_m,&\label{P12e} \\
& \; \bar{{\bf{U}}}_r \succeq 0,~ \bar{{\bf{U}}}_t \succeq 0. &\label{P12f}
\end{align}
\end{subequations}
In problem (P10), $\|\bar{{\bf{U}}}_k\|_*$ and $\widehat{\bar{{\bf{U}}}}_{k}^{n}$ are similarly defined as those in problem (P6). As such, an iteration algorithm similar to that for solving problem (P6) can be applied to obtain the optimal solution of Problem (P10). In a nutshell, due to involving the similar convex optimizations with respect to $\bar{{\bf{U}}}_r$ and $\bar{{\bf{U}}}_t$ as those for solving problem (P6), the complexity of the AO approach in the considered STAR-RIS-NOMA system is the same as that of Algorithm 2. The whole procedure of the AO approach is given in Algorithm 3.
\begin{algorithm}[t]
 %\textsl{}\setstretch{1.8}
% \renewcommand{\algorithmicrequire}{\textbf{Input:}}
% \renewcommand{\algorithmicensure}{\textbf{Output:}}
 \caption{AO Algorithm of Maximizing STAR-RIS-NOMA Covert Rate}
 \label{alg3}
 \begin{algorithmic}[1]
%     \REQUIRE System bound $B$, user information $P_a$, $\hat{R}_a$, $N_a$, $N_b$, $\sigma^2$, channel information $\ell_a$, $h_a$, $\ell_b$, $h_b$.
%  \ENSURE Power allocation $P_{b,1}$, $P_{b,2}$, rate splitting factor $\beta$, power allocation factor $\alpha$, and time phase $t_2$.
  \STATE Initialize the feasible ${\bf{\Theta}}_{r}^{0}$ and ${\bf{\Theta}}_{t}^{0}$.
  \STATE ${\bf{\Theta}}_{r}^{0}$, and ${\bf{\Theta}}_{t}^{0}$ can be equivalently transformed to $\bar{{\bf{U}}}_{r}^{0}$ and $\bar{{\bf{U}}}_{t}^{0}$.
  \STATE $\ell \gets 0$
  \REPEAT
  \STATE Given $\bar{{\bf{U}}}_{r}^{\ell}$ and $\bar{{\bf{U}}}_{t}^{\ell}$, solve the problem (P9) to obtain $\bar{ \alpha}_{1}^{\ell+1}$, and $\bar{\alpha}_{2}^{\ell+1}$.
  \STATE $n \gets 0$
  \REPEAT
  \STATE Given $\bar{ \alpha}_{1}^{\ell+1}$, and $\bar{\alpha}_{2}^{\ell+1}$ solve the problem (P10) to obtain $\bar{{\bf{U}}}_{r}^{n+1}$ and $\bar{{\bf{U}}}_{t}^{n+1}$.
  \STATE Update $n = n + 1$.
  \STATE $\bar{{\bf{U}}}_{r}^{0} \gets \bar{{\bf{U}}}_{r}^{n}$, $\bar{{\bf{U}}}_{t}^{0} \gets \bar{{\bf{U}}}_{t}^{n}$.
  \STATE $\epsilon \gets c_1 \epsilon$.
  \UNTIL $\sum_{k\in \{r, t\}} \left(\|\bar{{\bf{U}}}_k\|_{*}^{n} - \|\bar{{\bf{U}}}_k\|_{2}^{n} \right) \le \zeta_2$.
  \STATE $\bar{{\bf{U}}}_{r}^{0} \gets \bar{{\bf{U}}}_{r}^{\ell+1}$, $\bar{{\bf{U}}}_{t}^{0} \gets \bar{{\bf{U}}}_{t}^{\ell+1}$.
  \STATE Update $\ell = \ell + 1$.
  \UNTIL $R_{b}^{{\rm{NOMA}},\ell+1} - R_{b}^{{\rm{NOMA}},\ell} < \zeta_3$.
 \end{algorithmic}
\end{algorithm}

Compared with the STAR-RIS-RSMA scheme, the STAR-RIS-NOMA scheme optimizes only the two transmit power allocation factors $\bar{\alpha}_1$ and $\bar{\alpha}_2$, while the STAR-RIS-RSMA scheme optimizes the transmit power allocation factors $\alpha_0$, $\alpha_1$, and $\alpha_2$ and the rate allocation factors $\beta_1$ and $\beta_2$. Therefore, the STAR-RIS-RSMA scheme has a higher degree of freedom (DoF) than the STAR-RIS-NOMA scheme. With the increase of DoF, the STAR-RIS-RSMA scheme achieves a higher covert rate than the STAR-RIS-NOMA scheme, which will be verified by the simulation results.

Due to hardware impairments and imperfect CSI, interference cannot be completely eliminated during the SIC processing at the receiver. In this section, we consider the scenarios of imperfect SIC for the STAR-RIS-RSMA and STAR-RIS-NOMA schemes. Specifically, let $\omega$ denote the imperfect SIC coefficient, which satisfies $0 \le \omega \le 1$ \cite{9406993}. For the STAR-RIS-RSMA scheme with imperfect SIC, the achievable rates corresponding to the transmissions of the private streams of Bob and Grace can be expressed as: 
\begin{eqnarray}
\tilde{R}_{b,s_{1} (q)} = \log_2 \left(1 + \tfrac{\alpha_1 P_T |Z_{ab}|^2}{\omega \alpha_0 P_T |Z_{ab}|^2 + \alpha_2P_T|Z_{ab}|^2+ \sigma_{b}^{2}}\right).
\end{eqnarray}
and
\begin{eqnarray}
\tilde{R}_{g,s_{2} (q)} = \log_2 \left(1 + \tfrac{a_2 P_T |Z_{ag}|^2}{\omega \alpha_0 P_T |Z_{ag}|^2 + \alpha_1 P_T|Z_{ag}|^2 + \sigma_{g}^{2}}\right),
\end{eqnarray}
respectively. Then, the available rates of Bob and Grace under  imperfect SIC can be written as $R_b = \beta_1 R_{g,s_0 (q)} + \tilde{R}_{b,s_1(q)}$ and $R_g = \beta_2 R_{g,s_0 (q)} + \tilde{R}_{g,s_2(q)}$, respectively. 
Under imperfect SIC, the maximum covert rate of the STAR-RIS-RSMA can be obtained by absorbing $\omega$ in Algorithm 2, which is straightforward and the corresponding details are omitted here.  
For the STAR-RIS-NOMA scheme with imperfect SIC, the Bob' covert rate obtained from decoding private message $s_g (q)$ can be expressed as:
\begin{eqnarray}
\tilde{R}_{b,s_b (q)}^{^{\rm NOMA}} = \log_2 \left(1 + \frac{\bar{\alpha}_1 P_T |Z_{ab}|^2}{\omega \bar{\alpha}_2 P_T |Z_{ab}|^2 +   \sigma_{b}^{2}}\right).
\end{eqnarray}
Under imperfect SIC, the maximum covert rate of the STAR-RIS-NOMA scheme can be obtained by absorbing $\omega$ in Algorithm 3, which is also straightforward and the corresponding details are omitted. 

\begin{figure}[t]
    \begin{center}   
    \includegraphics[width=3in]{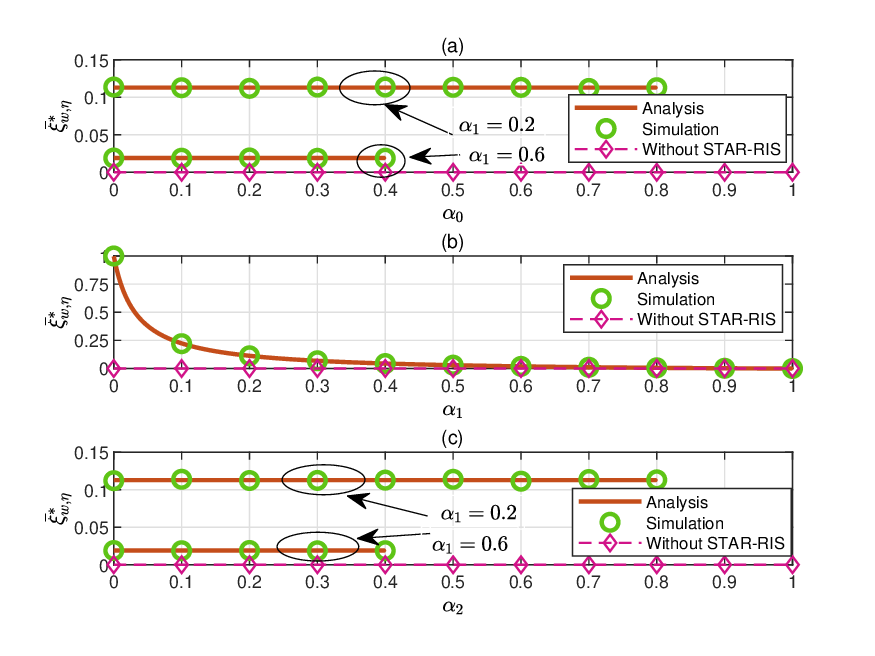}
		\caption{$\bar{\xi}_{w,\eta}^{*}$ versus transmit power allocation coefficients.}
		\label{fig2}
    \end{center}
    \vspace{-0.2in}
\end{figure}

\section{Simulation Results}

In this section, we present simulation results to verify the accuracy of the analytical expression of the MDEP and clarify the covert performance of the proposed optimization scheme. Throughout the simulations, the following system parameter setup is assumed unless otherwise stated. We consider a two-dimensional coordinate system, where Alice, Bob, Grace, Willie, and STAR-RIA are located at the positions (0, 0) m, (90, 0) m, (90, 10) m, (80, -5) m, and (80, 5) m, respectively \cite{9524501,9915477}. The system parameters related to the path-loss are set as $\chi_{ab} = \chi_{aw}= 3$, $\chi_{ar} = \chi_{rb}= \chi_{rg} = \chi_{rw} = 2$, $d_0=1$ m, $L_0 = 30$ dB, and $\lambda_{i} = 1$, $\forall i$ \cite{9496108,9834288}. Moreover, $\sigma_{w}^{2} = \sigma_{b}^{2} = \sigma_{g}^{2} = -90$ dBm, $K = 64$, imperfect SIC parameter $\omega = 0.01$, and the stopping thresholds for the iterations in the Algorithms are set as $\zeta_1 = \zeta_2 = \zeta_3 = 10^{-4}$.  

In Fig. \ref{fig2}, we investigate the MADEP versus Bob's transmit power allocation coefficients $\alpha_0$ and $\alpha_1$ with the fixed system parameters $P_{T} = 25$ dBm and $K_n = 40$.  As can be seen from Fig. \ref{fig2}(a), Fig. \ref{fig2}(b), and Fig. \ref{fig2}(c), our analysis results of $\bar{\xi}_{w,\eta}^{*}$ are consistent with the simulation results. In Figure \ref{fig2}(a), we fix $\alpha_1$ by setting $\alpha_1 \in \{ 0.2, 0.6 \} $ and vary the values of $\alpha_0$. From the curves in Fig.\ref{fig2}(a), we can see that $\bar{\xi}_{w,\eta}^{*}$ does not vary with respect to $\alpha_0$ when $\alpha_1$ is fixed. Similarly, $\bar{\xi}_{w,\eta}^{*}$ is a constant with respect to $\alpha_2$  when $\alpha_1$ is fixed. As discussed in Remark 5, $\bar{\xi}_{w,\eta}^{*}$ is only determined by $\alpha_1$, which is coincident with the curves in Fig. \ref{fig2}(a) and Fig. \ref{fig2}(c), respectively. From Fig. \ref{fig2}(b), we can see that $\bar{\xi}_{w,\eta}^{*}$ decreases monotonically with the decreasing $\alpha_1$, which indicates that Willie achieves a better detection performance with a higher transmit power for Bob's private message. As discussed in Remark 3, when $\alpha_1 \to 0$, Willie cannot make the binary decision correctly due to $\bar{\xi}_{w,\eta}^{*} \to 1$. On the contrary, when $\alpha_1 \to 1$, Willie can always make the correct binary decision due to $\bar{\xi}_{w,\eta}^{*} \to 0$. In such a case, allocating less transmit power to Bob's private message is beneficial to improve the covertness. Moreover, to maintain a certain non-zero level of covertness, Alice's transmit power cannot be solely allocated to Bob's transmission. In addition, considering the scenario without STAR-RIS, when Alice allocates additional transmit power for Bob's covert signal, Willie can easily detect the covert signal sent from Alice to Bob due to the absence of interference from STAR-RIS random signals. Therefore, in the scheme without STAR-RIS assistance, Willie's $\bar{\xi}_{w,\eta}^{*}$ is a zero.

\begin{figure}[t]
    \begin{center}   
    \includegraphics[width=3in]{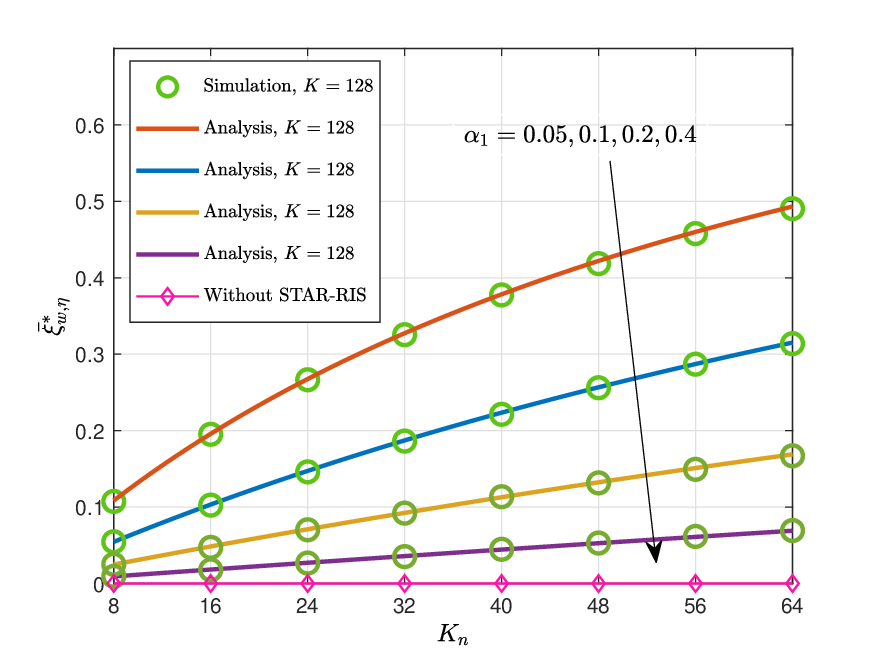}
		\caption{$\bar{\xi}_{w,\eta}^{*}$ versus $K_n$ values.}
		\label{fig3}
    \end{center}
    \vspace{-0.2in}
\end{figure} 

The relationship between the MADEP and the number of the reflecting elements $K_n$ is investigated in Fig. \ref{fig3}, where we set $P_T = 25$ dBm and $K=128$. The curves in Fig. \ref{fig3} show that $\bar{\xi}_{w,\eta}^{*}$ monotonically increases with the increasing $K_n$. The reason for this phenomenon is that an increasing in $K_n$ is helpful to increase more uncertainties to Willie's detection by involving more STAR-RIS reflecting elements. In addition, we can see from Fig. \ref{fig3} that a smaller $\alpha_1$ results in a larger $\bar{\xi}_{w,\eta}^{*}$, which is coincident with the results of Fig. \ref{fig2}. For the scheme without STAR-RIS, the Willie's $\bar{\xi}_{w,\eta}^{*}$ is zero.

\begin{figure}[t]
    \begin{center}   
    \includegraphics[width=3in]{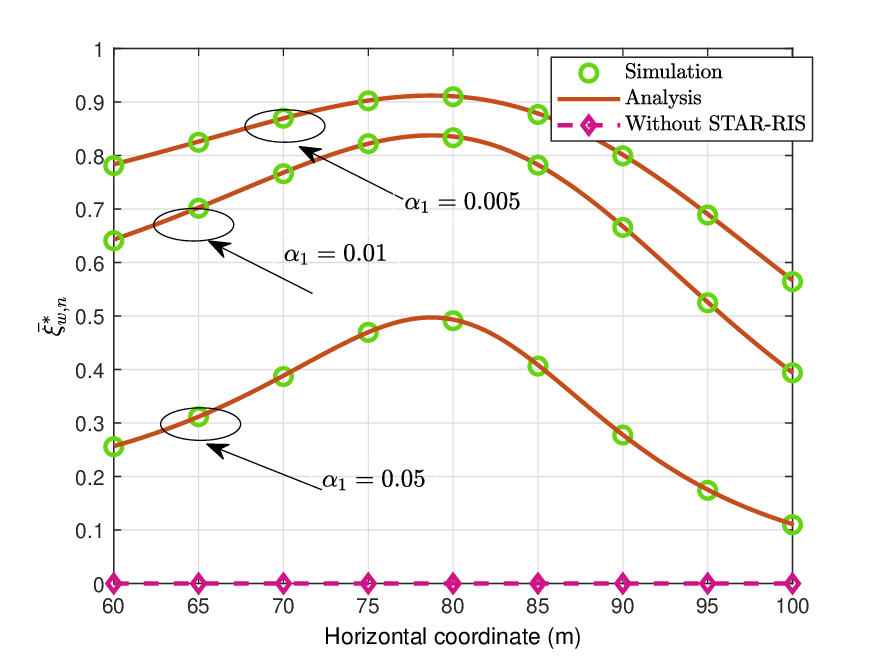}
		\caption{$\bar{\xi}_{w,\eta}^{*}$ versus STAR-RIS value of Horizontal coordinate.}
		\label{fig40}
    \end{center}
    \vspace{-0.2in}
\end{figure}

In Fig. \ref{fig40}, the relationship between Willie's MADEP and the location of STAR-RIS is revealed. For the simulations corresponding to the results of Fig. \ref{fig40}, we set $K = 128$, $k_N = K_m = 64$ and $P_T = 25$ dBm, and the vertical coordinate of STAR-RIS is set to be 5 m. From Fig. \ref{fig40}, We can observe that as the STAR-RIS horizontal coordinate increases, $\bar{\xi}_{w,\eta}^{*}$ first increases monotonically, and then decreases monotonically after reaching a maximum value. The reason for this phenomenon is that as the STAR-RIS approaches Willie, the uncertainty observed by Willie is intensified by the STAR-RIS reflections, which in turn decreases Willie's detection performance. Willie achieves the highest  MADEP when the position of the STAR-RIS moves to (80, 5) m. As the STAR-RIS moves away from the BS and Willie, Willie's MADEP decreases rapidly due to $\frac{\partial \bar{\xi}_{w,\eta}^{*}}{\partial \varphi} < 0$, where $\varphi = \tfrac{L_{ar} L_{rw}}{L_{aw}}$. When the STAR-RIS is located far away from Alice and Willie, the signal received by the STAR-RIS from Alice is weakened and the uncertainty resulted by the STAR-RIS reflections to Willie is reduced. Thus, the MADEP of Willie is decreased, which can be clarified by the expression $\varphi = \tfrac{L_{ar} L_{rw}}{L_{aw}}$. In order to minimize Willie's detection performance, the STAR-RIS needs to be positioned close to Willie to result in uncertainty. However, the covert rate of Bob is not only determined by the covert constraint, but also affected by Grace's QoS requirements and channel gains. In addition, Willie's MADEP can be further increased by decreasing $\alpha_1$, which may decrease Bob's covert rate. Thus, optimizing the transmit power allocation is necessary to improve the covert rate. As expected, the RSMA without STAR-IRS scheme always have a zero $\bar{\xi}_{w,\eta}^{*}$.

\begin{figure}[t]
    \begin{center}   
    \includegraphics[width=3in]{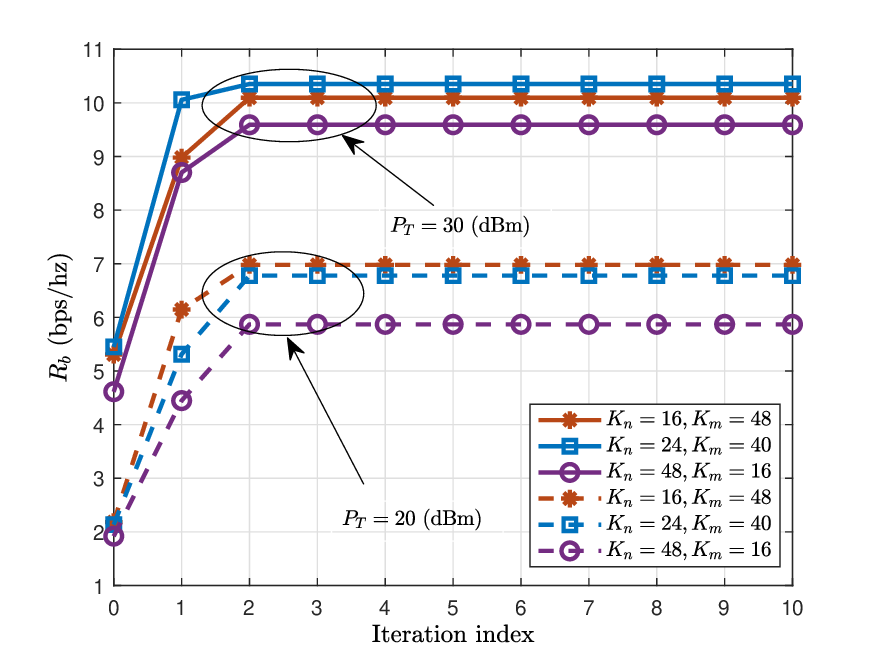}
		\caption{Convergence performance of the proposed AO algorithm.}
		\label{fig4}
    \end{center}
    \vspace{-0.2in}
\end{figure}

\begin{figure}[t]
    \begin{center}   
    \includegraphics[width=3in]{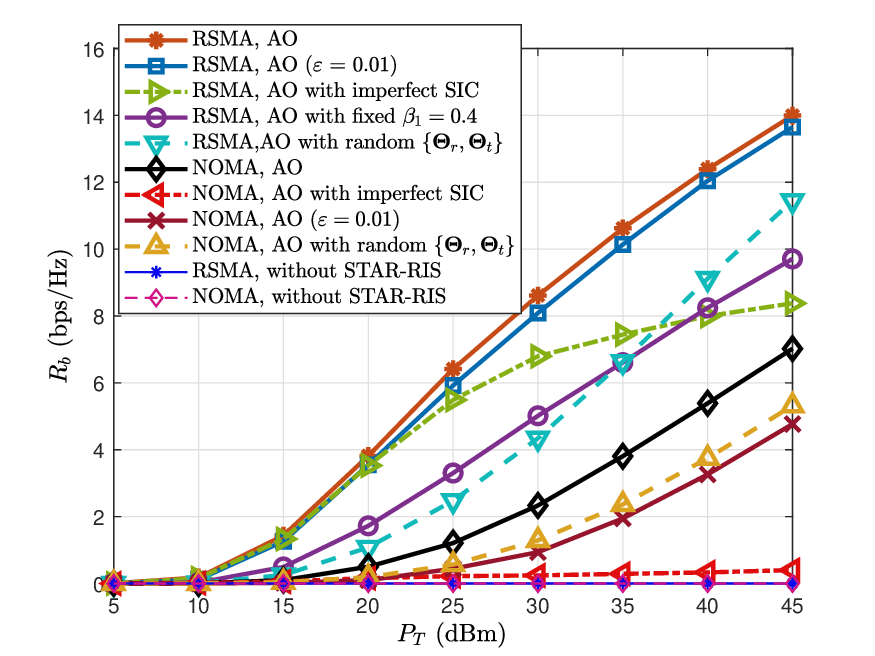}
		\caption{Covert rate versus transmit power.}
		\label{fig5}
    \end{center}
    \vspace{-0.2in}
\end{figure}

The convergence performance of the proposed AO algorithm in depicted in Fig. \ref{fig4}, where we set $K = 64$ and $P_T \in \{20, 30\}$ dBm. In Fig. \ref{fig4}, the values of the objective function $R_b$ at the iteration index 0 are obtained with randomly generated initial parameters. From Fig. \ref{fig4}, it can be observed that the optimal solutions for all the settings are achieved after three iterations, which indicates the proposed AO algorithm has a fast convergence speed. Moreover, the curves in Fig. \ref{fig4} demonstrate that the stable values of $R_b$ after three iterations, which verifies that the proposed AO algorithm has a stable convergence performance. Since the convergence speed is fast, the proposed AO algorithm has a low implementation complexity due to the less iterations required to obtain the optimal solution.
In addition, we can observe from Fig. \ref{fig4} that Bob's covert rate can be effectively improved by increasing the transmit  power and we will discuss the relationship between transmitting power and covert rate in details in Fig. \ref{fig5}. With the transmit power $P_T = 30$ dBm and $K_n = 16$, the covert rate of Bob can be improved by increasing the number of the reflective elements $K_n$ of the STAR-RIS. However, by increasing $K_n$ further, the covert rate of Bob decreases. Moreover, With the transmit power $P_T = 20$ dBm and $K_n = 16$, the increasing $K_n$ decreases the covert rate of Bob. The reason for this phenomenon is that in the low transmit power region, the STAR-RIS needs to allocate more elements for the transmissions and Alice needs to allocate more transmit power and more common rate to Grace in order to meet Grace's QoS requirements.  

In Fig. \ref{fig5}, we illustrate the relationship between the covert rate and the transmit power. The corresponding simulation parameters are set as follows: $R_{g}^{\min} = 1$ bps/Hz, $K_n=32$, $K_m=32$, and $\varepsilon = 0.05$. From Fig. \ref{fig5}, we have the following observations: The covert rate achieved by all schemes increases with the increasing of the transmit power except for the scheme without STAR-RIS. The proposed STAR-RIS-RSMA scheme with the AO algorithm outperforms other schemes in the terms of the covert rate. For the scheme without STAR-RIS, the covert rate obtained is always zero, since covert communication cannot be realized.
Since the STAR-RIS-RSMA scheme can jointly optimize the transmit power and common rate allocations, a higher DoF can be achieved by the STAR-RIS-RSMA scheme compared to the STAR-RIS-NOMA scheme, which can only optimize the power allocation factors $\bar{\alpha}_1$ and $\bar{\alpha}_2$. Therefore, the STAR-RIS-RSMA scheme significantly improves the covert rate. The proposed STAR-RIS-RSMA scheme achieves a transmit power gain of about 8 dBm at the covert rate level of $R_b = 10$ bps/Hz compared to the random phase shift scheme. 
In addition, the difference between the covert rates achieved by the STAR-RIS-RSMA and STAR-RIS-NOMA schemes gradually increases as the transmit power increases, which further verifies that the STAR-RIS-RSMA scheme is superior to the STAR-RIS-NOMA scheme in the terms of covert rate.
For the STAR-RIS-RSMA scheme, the optimal common rate allocation achieves a  higher covert rate than the fixed rate allocation scheme. 
For example, at the covert rate level of $R_b = 10$ bps/Hz, the STAR-RIS-RSMA scheme with the common rate allocation optimization almost achieves a transmit power gain of 10 dBm compared to the STAR-RIS-RSMA scheme with the fixed rate allocation. 
Furthermore, in the low transmit power region, the STAR-RIS-RSMA scheme with the fixed rate allocation achieves a higher $R_b$ compared to the STAR-RIS-RSMA with the random reflection and transmission phase shifts. However, in the high transmit power region, the STAR-RIS-RSMA scheme with the random STAR-RIS reflection and transmission phase shifts obtains a higher $R_b$. Therefore, the results in Fig. \ref{fig5} validate the importance of the optimized common rate allocation. Similarly, for the fixed transmit power allocation, the obtained $R_b$ is smaller than that achieved by the optimal transmit power allocation decreased too(not depicted in Fig. \ref{fig5} due to the crowded space).
In addition, the schemes with the STAR-RIS reflection and transmission beamforming optimization achieve the  higher covert rates compared to the schemes with the random reflection and transmission phase shifts. 
As such, the STAR-RIS-RSMA scheme with the optimized transmit power and common rate allocations achieves a higher cover rate than the that achieved by the STAR-RIS-NOMA scheme using the AO algorithm,  which indicates that jointly optimized transmit power and common rate allocations can significantly improve the covert rate.
With the increasing of the covert constraint level (i.e., decreasing $\varepsilon$), the covert rate achieved by all the schemes decrease due to the fact that a higher covert constraint levels is more stringent, which limits Bob's received signal power and results in a lower covert rate.
On the other hand, the effect of imperfect SIC on the covert rate is obvious, i.e., all the schemes with imperfect SIC obtain a lower covert rate compared to the perfect SIC counterparts. However, with imperfect SIC, the STAR-RIS-RSMA scheme can still obtain a higher covert rate than that of the STAR-RIS-NOMA scheme by flexibly adjusting the transmit power allocation and rate allocation factors.  

\begin{figure}[t]
    \begin{center}   
    \includegraphics[width=3in]{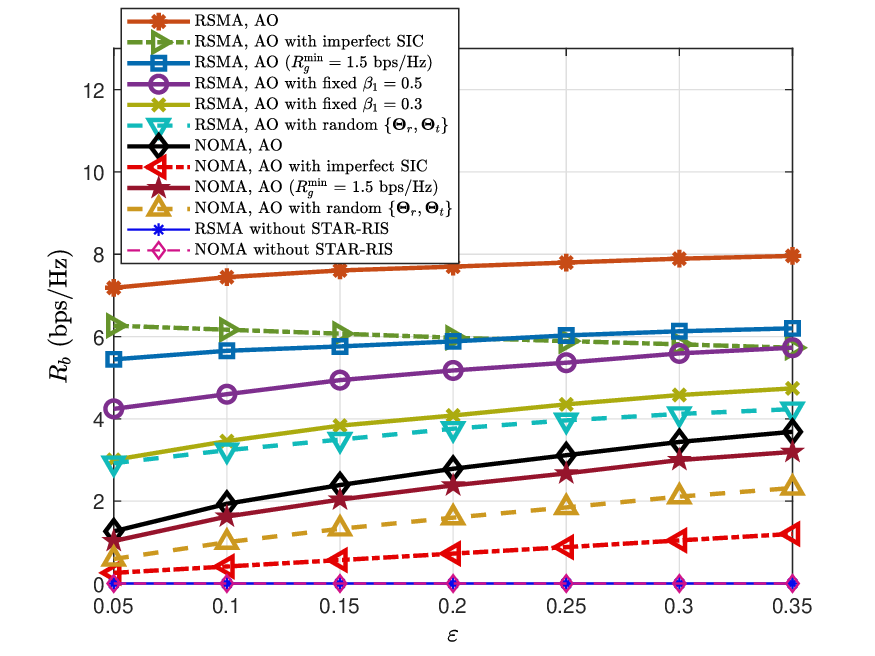}
		\caption{Covert rate versus $\varepsilon$.}
		\label{fig6}
    \end{center}
    \vspace{-0.2in}
\end{figure}

In Fig. \ref{fig6}, the relationship between the covert rate and the covert parameter $\varepsilon$ is depicted. The corresponding simulation parameters are set to $P_T = 25$ dBm, $K_n=32$,  $K_m=32$,  and $R_{g}^{\min} = 0.5$. It can be observed from Fig. \ref{fig6} that the covert rates achieved by all the schemes increase with the increasing of $\varepsilon$ except for that of the STAR-RIS-RSMA scheme with imperfect SIC and without STAR-RIS scheme. This is because the covert constraint on Willie relaxes as $\varepsilon$ increases, which allows Alice to allocate more transmit power to Bob's private message as verified by Equation \eqref{MADEP_2}. 
The covert rate achieved by the STAR-RIS-RSMA scheme with imperfect SCI decreases slowly with increasing $\varepsilon$. The covert rate achieved by the STAR-RIS-RSMA scheme with imperfect SIC is mainly dependent on $R_{b}^{c}$ compared to the case of perfect SIC. For the scenario without STAR-RIS, the covert rate achieved by Bob is always zero. Similarly, in the subsequent simulations, the covert rates achieved by Bob are not discussed in detail since they are all zero. In addition, the STAR-RIS-NOMA scheme with imperfect SIC achieves a much lower covert rate than other schemes except for the scheme without STAR-RIS. 
In Fig. \ref{fig6}, the proposed STAR-RIS-RSMA scheme with the AO algorithm achieves a significantly higher covert rate than the STAR-RIS-NOMA scheme in the low $\varepsilon$ region. Compared to the STAR-RIS-NOMA scheme which optimizes only $\bar{\alpha}_1$ and $\bar{\alpha}_2$, the STAR-RIS-RSMA scheme can optimize not only the transmit power allocation factors, but also the common rate allocation factors $\beta_1$ and $\beta_2$ to obtain more DoF.
Moreover, for different fixed common rate allocation schemes, it is clear that a higher $\beta_1$ leads to a higher $R_b$. This is because the objective function $R_b$ is composed of the rates of the common and private streams and the covert rate can be increased by increasing $\beta_1$. However, the STAR-RIS-RSMA scheme  with the AO algorithm still achieves the highest covert rate, which once again proves the importance of optimal common rate allocation.
Furthermore, the covert rate $R_b$ of the STAR-RIS-NOMA scheme increases rapidly with increasing $\varepsilon$. Nevertheless, the covert rate $R_b$ achieved by the STAR-RIS-RSMA scheme with the AO algorithm is still higher than that of the STAR-RIS-NOMA scheme with the AO algorithm. For example, at the covertness level of $\varepsilon = 0.05$, the covert rate achieved by the STAR-RIS-RSMA scheme with the AO algorithm is about 6 bps/Hz greater than that achieved by the STAR-RIS-NOMA scheme with the AO algorithm. In addition, in the high $\varepsilon$ region, for example, $\varepsilon = 0.35$, the STAR-RIS-RSMA scheme achieves a covert rate that is still about 4 bps/Hz higher than the STAR-RIS-NOMA scheme. Similarly, the schemes with the optimized STAR-RIS reflection and transmission beamforming can significantly improve the covert performance. For example, at the level of $\varepsilon = 0.1$, the $R_b$ achieved by the STAR-RIS-RSMA scheme with the beamforming optimization is approximately 4.2 bps/Hz higher than that of the STAR-RIS-RSMA scheme with the random phase shifts. In the same case, the STAR-RIS-NOMA scheme with the beamforming optimization can achieve a gain of about 1 bps/Hz compared to the STAR-RIS-NOMA scheme with the random phase shifts.
By increasing Grace's QoS requirements, i.e., increasing $R_{g}^{\rm{min}}$, the covert rate achieved by Bob decreases. This is because Alice needs to allocate more transmit power to $\alpha_0$ or $\alpha_2$ to meet Grace's QoS requirements.

\begin{figure}[t]
    \begin{center}   
    \includegraphics[width=3in]{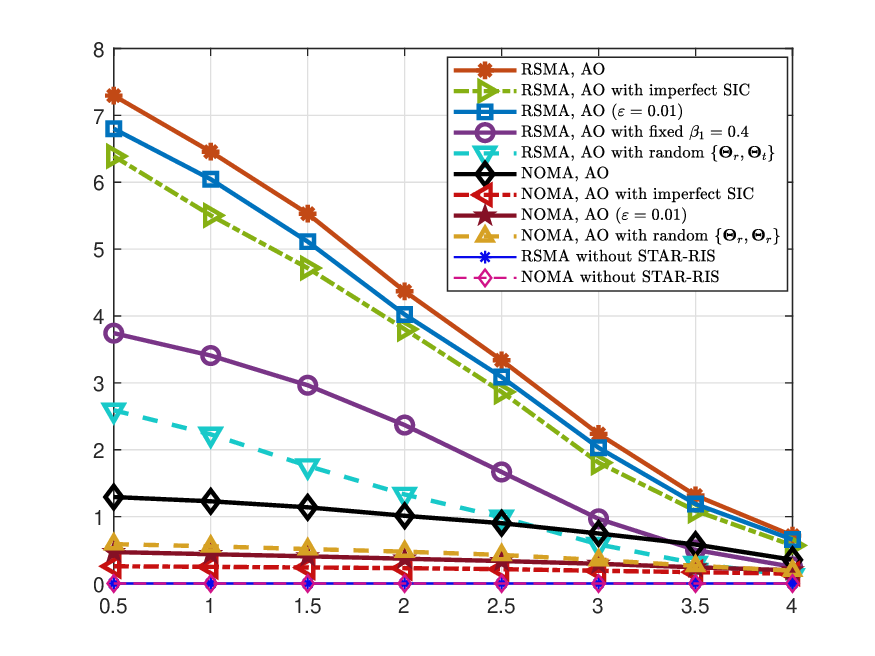}
		\caption{Covert rate versus $R_{g}^{\min}$.}
		\label{fig7}
    \end{center}
    \vspace{-0.2in}
\end{figure}

Fig. \ref{fig7} depicts the relationship between the covert rate and Grace's target rate $R_{g}^{\rm{min}}$. In Fig. \ref{fig7}, the simulation parameters are set as: $P_T = 25$ dBm, $K_n=32$,  $K_m=32$, and $\varepsilon=0.05$. First, we can observe that the covert rates achieved by all the schemes decrease with increasing $R_g^{\min}$. This is because as $R_g^{\min}$ increases, Grace's QoS requirements become more stringent, which leads to a decreasing in the covert rate achieved by Bob. Moreover, the covert rate achieved by the proposed STAR-RIS-RSMA scheme with the AO algorithm decreases rapidly with increasing $R_g^{\min}$. This is because as $R_g^{\min}$ increases, to ensure Grace's QoS requirements, Alice needs to allocate more transmit power not only to the common streams and Grace's private stream, but also to the common rate to Grace. However, the STAR-RIS-RSMA scheme still outperforms the benchmark scheme even in the high $R_g^{\min}$ region.
For example, at the QoS requirement level of $R_{g}^{\min} = 3$ bps/Hz, the covert rate $R_b$ achieved by the STAR-RIS-RSMA scheme with the AO algorithm is about 1.4 bps/Hz larger than that of the STAR-RIS-NOMA scheme with the AO algorithm. The covert rate $R_b$ achieved by the STAR-RIS-RSMA scheme with the random phase shifts is about 0.85 bps/Hz larger than that of the STAR-RIS-NOMA scheme with the random phase shifts.
Also, the STAR-RIS-RSMA scheme with the fixed common rate allocation achieves a higher $R_b$ in the low and middle $R_g^{\min}$ regions than the STAR-RIS-NOMA scheme with the AO algorithm, while in the high $R_g^{\min}$ region, the opposite result is obtained. This is because, in the high $R_g^{\min}$ region, the fixed common rate allocation scheme results in Grace's QoS requirements not being met, which leads to a lower $R_b$. This result also validates the importance of optimizing the rate allocation. In the low $R_{g}^{\min}$ region, optimizing the STAR-RIS reflection and transmission beamforming significantly increases the covert rate. For example, at the level of $R_{g}^{\min} = 1$ bps/Hz, compared with the STAR-RIS-RSMA scheme with the random phase shifts, the covert rate can be improved by about 2.5 bps/Hz by the STAR-RIS-RSMA scheme with the reflection and transmission beamforming. Even in the high $R_{g}^{\min}$ region, the covert rate achieved by the STAR-RIS-RSMA scheme with the reflection and transmission beamforming optimization is still higher than that of the STAR-RIS-RSMA scheme with the random phase shifts.
Furthermore, we find that the STAR-RIS-RSMA scheme with imperfect SIC achieves a higher covert rate than the STAR-RIS-NOMA scheme with imperfect SIC, even in the high $R_{g}^{{\rm{min}}}$ region.

\begin{figure}[t]
    \begin{center}   
    \includegraphics[width=3in]{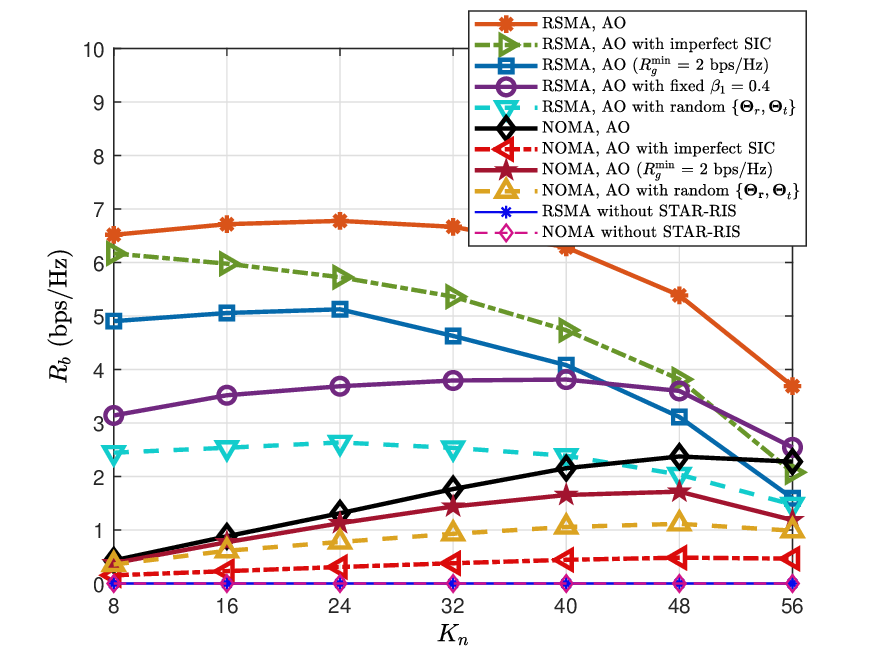}
		\caption{The impacts of $K_n$ on covert rate.}
		\label{fig8}
    \end{center}
    \vspace{-0.2in}
\end{figure}

In Fig. \ref{fig8}, the impacts of the number of the reflecting elements $K_n$ on the covert rate are revealed. In the simulations, we set $K = 64$, $P_T = 25$ dBm, $\varepsilon = 0.1$, and $R_{g}^{\min} = 1$. The results in Fig. \ref{fig8} show that the covert rates achieved by all the schemes first increase with the increasing $K_n$ except for the STAR-RIS-RSMA scheme with imperfect SIC. After $K_n$ approaches a certain number, the covert rates achieved by all schemes then decrease with the increasing $K_n$. Thus, a trade-off exists between $K_n$ and $K_m$ to achieve the allowed maximum covert rate. Since $\bar{\xi}_{w,\eta}^{*}$ increases with the increasing $K_n$, an increased $K_n$ looses the covertness constraint accordingly. Moreover, an increased $K_n$ brings more reflection power to enhance the equivalent channel gain from Alice to Bob. Thus, the covert rate first increases with the increasing $K_n$. On the other hand, since $K_m$ decreases accordingly with the increasing $K_n$ and a decreased $K_m$ requires to allocate more transmit power to guarantee the QoS requirements of Grace due to the weakened channel gain from Alice to Grace, a small $K_m$ (equivalently a large $K_n$) results in less transmit power to transmit Bob's message and the covert rate decreases accordingly. However, the proposed STAR-RIS-RSMA scheme with the AO algorithm achieves the highest covert rate among all the schemes in the considered whole $K_n$ region.
In Fig. \ref{fig8}, we find that the covert rate achieved by the STAR-RIS-RSMA scheme with imperfect SIC decreases as $K_n$ increases. The reason for this phenomenon is that the private stream rate achieved by the STAR-RIS-RSMA scheme with imperfect SIC is significantly reduced, while the rate achieved by decoding the common stream is not affected by $\omega$. Therefore, increasing $K_m$ can increase the covert rate of the STAR-RIS-RSMA scheme with imperfect SIC. For the STAR-RIS-NOMA scheme with imperfect SIC, Bob suffers severe interference in decoding private messages due to imperfect SIC, which results in a low covert rate. To overcome the detrimental effect of imperfect SIC on the covert rate, more STAR-RIS elements should be allocated to the reflective end to increase Bob's covert rate. Furthermore, we can observe that the STAR-RIS-RSMA scheme with the AO algorithm obtains the maximum covert rate at $K_n = 32$, while the STAR-RIS-NOMA scheme with the AO algorithm obtains the maximum covert rate at $K_n = 48$. This suggests that for the STAR-RIS-RSMA scheme, the reflecting and transmitting elements of STAR-RIS should be evenly distributed in order to increase Bob's covert rate, while for the STAR-RIS-NOMA scheme, more elements should be allocated to the reflecting end. When only part of the system parameters are optimized, the obtained covert performance decreases. For example, in the region of $K_n = 32$, the Rb obtained by the STAR-RIS-RSMA scheme with a random phase shift scheme is 3.5 bps/Hz lower compared to the STAR-RIS-RSMA scheme with the AO algorithm.

\begin{figure}[t]
    \begin{center}   
    \includegraphics[width=3in]{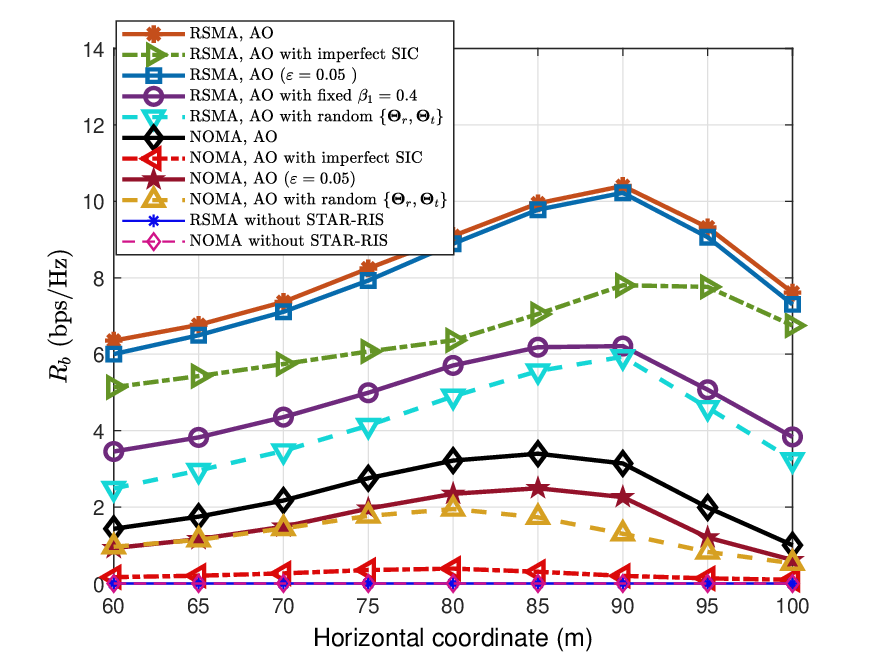}
		\caption{The impacts of STAR-RIS location on covert rate.}
		\label{fig10}
    \end{center}
    \vspace{-0.2in}
\end{figure}

Fig. \ref{fig10} investigates the relationship between the covert rate and location of the STAR-RIS. The simulation parameters in Fig. \ref{fig10} are set as: $P_T = 30$ dBm, $K_n = K_m = 32$, $\varepsilon = 0.1$, and $R_{g}^{\min} = 1$ bps/Hz. Moreover, the vertical coordinate of the STAR-RIS location is fixed in the simulation, while only the horizontal coordinate of the STAR-RIS location varies in the simulation. In Fig. \ref{fig10}, the covert rates achieved by all the schemes first increase as the STAR-RIS moves horizontally far away from Alice. Then, the covert rates decrease by continually moving the STAR-RIS horizontally far away from Alice. The reason for this phenomenon is that as the STAR-RIS gets closer to Willie, the uncertainty observed by Willie is intensified by the STAR-RIS reflections, which results in a decreasing of Willie's detection performance, such that Alice is allowed to allocate more transmit power to Bob's private stream. 
On the other hand, as the STAR-RIS gradually approaches Bob and Grace, the channel gains of Bob and Grace are further enhanced, which makes Grace's QoS requirements more easier to be met and thus improves the covert rate. An interesting phenomenon is that for the STAR-RIS-RSMA scheme, the highest covert rate can be achieved when the STAR-RIS is located close to Bob and Grace. On the other hand, the STAR-RIS-NOMA scheme with the AO algorithm obtained the highest covert rate when the STAR-RIS is located between Willie and Bob. This verifies that the STAR-RIS-RSMA scheme with the optimizations of the transmit power allocation and the common rate allocation can achieve a higher covert performance than the STAR-RIS-NOMA scheme with the transmit power optimization only. Based on the above discussions, for different schemes, the appropriate location of the STAR-RIS can be selected to maximize the covert rate of Bob.

\section{Conclusions}

In this paper, covert communication in STAR-RIS-RSMA systems has been investigated with the aim to maximize the covert rate. Closed-form expression for the MADEP has been derived and adopted as a metric to measure the covert performance of the considered system. To achieve the maximum covert rate, the system parameters, including the transmit power allocation, common rate allocation, and STAR-RIS reflection and transmission beamforming have been jointly considered in the optimization. Due to the non-convex of the formulated covert rate maximization problem and the highly coupled system parameters, an AO algorithm has been designed to obtain the optimized solution with the aid of decoupling the original problem into three sub-problems of the transmit power allocation, common rate allocation, and STAR-RIS reflection and transmission beamforming, respectively. The optimal solution of the transmit power allocation and common rate allocation for the corresponding sub-problems have been derived. Moreover, a PSCA
scheme to obtain the optimized STAR-RIS reflection and transmission beamforming has also been. Simulation results have verified the superior covert performance achieved by the proposed AO algorithm. The superiority of our proposed scheme over the benchmark schemes has been clearly highlighted.

\section*{Appendix A: Proof of Theorem 1}
\setcounter{equation}{0}

\renewcommand{\theequation}{A.\arabic{equation}}

According to the expression for DEP in \eqref{w_n}, the optimal detection threshold can be characterized in the following three cases:

Case I: $\eta < \sigma_{w}^{2} + \delta_{1} |h_{aw}|^2$. In this case, we have $\xi_{w,\eta} = 1$, so that the optimal detection threshold does not exist.  

Case II: $\sigma_{w}^{2} + \delta_1 |h_{aw}|^2 \le \eta \le \sigma_{w}^{2} + \delta_2 |h_{aw}|^2$. In this case, since $\xi_{w,\eta}$ is a monotonically decreasing function with respect to $\eta$, the optimal detection threshold is determined  by $\eta^* = \sigma_{w}^{2} + \delta_2 |h_{aw}|^2$.  

Case III: $\eta > \sigma_{w}^{2} + \delta_2 |h_{aw}|^2$. In this case, the first derivative of $\xi_{w,\eta}$ with respect to $\eta$ can be derived as:
\begin{flalign}
\frac{\partial \xi_{w,\eta}}{\partial \eta } = \frac{e^{\frac{\delta_2 |h_{aw}|^2 + \sigma_{w}^{2} - \eta}{\delta_2 K_n \varphi^{-1}}}}{\delta_2 K_n \varphi^{-1}} - \frac{e^{\frac{\delta_1 |h_{aw}|^2 + \sigma_{w}^{2} - \eta}{\delta_1 K_n \varphi^{-1}}}}{\delta_1 K_n \varphi^{-1}}. \label{piandao}
\end{flalign}
Then, let $\frac{\partial \xi_{w,\eta}}{\partial \eta} = 0$, the optimal solution in this case can be obtained as:
\begin{eqnarray}
\eta^{\dag} = \sigma_{w}^{2} + \frac{\delta_1 K_n}{\alpha_1 \varphi} \ln \left(\frac{1}{\alpha_0 + a_2}\right). \label{jizhi}
\end{eqnarray}

Based on \eqref{piandao} and \eqref{jizhi}, the monotonicity of $\xi_{w, \eta}$ with respect to $\eta$ is discussed as follows: 1) when $\eta^{\dag} \ge \sigma_{w}^{2} + \delta_2 |h_{aw}|^2$, we have $\frac{\partial \xi_{w,\eta}}{\partial \eta} < 0$ if $\eta \in (\sigma_{w}^{2} + \delta_2 |h_{aw}|^2,~ \eta^{\dag})$  and  $\frac{\partial \xi_{w,\eta}}{\partial \eta} > 0$ 
if  $\eta \in (\eta^{\dag},~ + \infty)$. Thus, $\xi_{w,\eta}$ first decreases monotonically with the increasing $\eta$ and then increases monotonically with the increasing $\eta$ once $\eta > \eta^{\dag}$. As a result, the optimal detection threshold for minimizing $\xi_{w,\eta}$ is given by $\eta^* = \eta^{\dag}$. 2) when $\eta^{\dag} < \sigma_{w}^{2} + \delta_2 |h_{aw}|^2$, we have $\frac{\partial \xi_{w, \eta}}{\partial \eta} > 0$ if $\eta \in \left(\sigma_{w}^{2} + \delta_1 |h_{aw}|^2,~ +\infty \right)$. Thus, $\xi_{w, \eta}$ is a monotonically increasing function with respect to $\eta$. Consequently, the optimal solution is given by $\eta^* = \sigma_{w}^{2} + \delta_2 |h_{aw}|^2$. Based on the above results derived for the three cases, the optimal detection threshold can be written as \eqref{n_1}. By substituting \eqref{n_1} into \eqref{w_n}, the minimum DEP is derived as \eqref{w_n1}.

\section*{Appendix B: Proof of Proposition 1}
\setcounter{equation}{0}

\renewcommand{\theequation}{B.\arabic{equation}}
We use the contradiction to prove proposition 1. Assume that the maximum value of the objective function of problem (P2) can be achieved when $\alpha_0 + \alpha_1 + \alpha_2 < 1$ and rewrite $R_b$ as: 
\begin{eqnarray}
R_b = \beta_1 \log_2 \left(1 + \frac{\alpha_0 P_{T} |Z_{ag}|^2}{(1-\alpha_0) P_{T}|Z_{ag}|^2 + \sigma_{g}^2} \right) \nonumber \\+ \log_2 \left(1 + \frac{\alpha_1 P_{T} |Z_{ab}|^2}{\alpha_2 P_{T}|Z_{ab}|^2 +\sigma_{b}^2}\right).~~~~~~
\end{eqnarray}
It is readily to show that $R_b$ is a monotonically increasing function with respect to $\alpha_1$ by fixing $\alpha_0$. By taking the first derivative of $\bar{\xi}_{w,\eta}^{*}$ with respect to $\alpha_1$, it can be proven that  $\frac{\partial \bar{\xi}_{w,\eta}^{*}}{\partial \alpha_1} < 0$, which means 
that a certain covertness requires $\alpha$ to be no large than a threshold. 
Based on the above discussions, for any fixed $\alpha_0$ and $\alpha_2$, we can scale up $\alpha_1$ by a factor of $c_2$ ($c_2 > 1$), such that $\alpha_0 + \alpha_1 +\alpha_2 = 1$ holds, which results in an increased objective function value due to the increasing monotonicity of $R_b$ with respect to $\alpha_1$. This contradicts the original assumption of $\alpha_0 + \alpha_1 +\alpha_2 < 1$ to achieve the maximum objective function value. Also, $\alpha_1$ is  constrained by the monotonicity of $\bar{\xi}_{w,\eta}^{*}$, which limits the increasing of the objective function value and thus complete the proof. 

\section*{Appendix C: Proof of Proposition 2}
\setcounter{equation}{0}

\renewcommand{\theequation}{C.\arabic{equation}}

As shown in step 5 in Algorithm 2, the optimal solution $\alpha_{0}^{\ell+1},~ \alpha_{1}^{\ell+1}$, and $\alpha_{2}^{\ell+1}$ can be computed with the given $\beta_{1}^{\ell},~ \beta_{2}^{\ell}$, ${\bf{U}}_{r}^{\ell}$, and ${\bf{U}}_{t}^{\ell}$. Thus, we have the following inequality
\begin{eqnarray}
 R_b \left(\beta_{1}^{\ell}, \beta_{2}^{\ell}, {\bf{U}}_{r}^{\ell}, {\bf{U}}_{t}^{\ell}, \alpha_{0}^{\ell}, \alpha_{1}^{\ell}, \alpha_{2}^{\ell} \right)~~~~~~~~~~~~~~~~~~ \nonumber \\ 
\le R_b \left(\beta_{1}^{\ell}, \beta_{2}^{\ell}, {\bf{U}}_{r}^{\ell}, {\bf{U}}_{t}^{\ell},  \alpha_{0}^{\ell+1}, \alpha_{1}^{\ell+1}, \alpha_{2}^{\ell+1} \right). \label{t1}
\end{eqnarray}
Then, as shown in step 6 in Algorithm 2, the optimal solution $\beta_{1}^{\ell+1}$ and $\beta_{2}^{\ell+1}$ can be computed with the given $\alpha_{0}^{\ell+1}, \alpha_{1}^{\ell+1}, \alpha_{2}^{\ell+1}$, ${\bf{U}}_{r}^{\ell}$, and ${\bf{U}}_{t}^{\ell}$. Thus, the inequality \eqref{t1} is updated as
\begin{eqnarray}
R_b \left(\beta_{1}^{\ell}, \beta_{2}^{\ell}, {\bf{U}}_{r}^{\ell}, {\bf{U}}_{t}^{\ell}, \alpha_{0}^{\ell+1}, \alpha_{1}^{\ell+1}, \alpha_{2}^{\ell+1} \right)~~~~~~~~~~~~~~~~ \nonumber \\ \le
R_b \left(\beta_{1}^{\ell+1}, \beta_{2}^{\ell+1}, {\bf{U}}_{r}^{\ell}, {\bf{U}}_{t}^{\ell}, \alpha_{0}^{\ell+1}, \alpha_{1}^{\ell+1}, \alpha_{2}^{\ell+1}\right). \label{t2}
\end{eqnarray}
Furthermore, as shown in step 14 in Algorithm 2, the optimal solution ${\bf{U}}_{r}^{\ell+1}$ and ${\bf{U}}_{t}^{\ell+1}$ can be computed with the given $\alpha_{0}^{\ell+1}, \alpha_{1}^{\ell+1}, \alpha_{2}^{\ell+1}, \beta_{1}^{\ell+1}$, and $\beta_{2}^{\ell+1}$ and we have the inequality as follows:
\begin{eqnarray}
R_b \left(\beta_{1}^{\ell+1}, \beta_{2}^{\ell+1}, {\bf{U}}_{r}^{\ell}, {\bf{U}}_{t}^{\ell}, \alpha_{0}^{\ell+1}, \alpha_{1}^{\ell+1}, \alpha_{2}^{\ell+1}\right)~~~~~~~~~~~~~~~~ \nonumber \\ \le
R_b\left(\beta_{1}^{\ell+1}, \beta_{2}^{\ell+1}, {\bf{U}}_{r}^{\ell+1}, {\bf{U}}_{t}^{\ell+1}, \alpha_{0}^{\ell+1}, \alpha_{1}^{\ell+1}, \alpha_{2}^{\ell+1}\right).~  \label{t3}
\end{eqnarray}
Based on the above inequalities, we can readily arrive at
\begin{eqnarray}
R_b \left(\beta_{1}^{\ell}, \beta_{2}^{\ell}, {\bf{U}}_{r}^{\ell}, {\bf{U}}_{t}^{\ell}, \alpha_{0}^{\ell}, \alpha_{1}^{\ell}, \alpha_{2}^{\ell} \right)~~~~~~~~~~~~~~~~~~~~~~~~~~~~ \nonumber \\
\le R_b \left(\beta_{1}^{\ell+1}, \beta_{2}^{\ell+1}, {\bf{U}}_{r}^{\ell+1}, {\bf{U}}_{t}^{\ell+1},  \alpha_{0}^{\ell+1}, \alpha_{1}^{\ell+1}, \alpha_{2}^{\ell+1} \right).  \label{t4}
\end{eqnarray}
According to \eqref{t1}, \eqref{t2}, \eqref{t3}, and \eqref{t4}, it can be shown that the objective function value of problem (P1) is progressively increasing over the iterations. Moreover, the objective function $R_b$ of problem (P1) clearly has an upper bound for any choice of feasible ${\bf{\Theta}}_r$ and ${\bf{\Theta}}_t$ and $\alpha_0$, $\alpha_1$, $\alpha_2$, $\beta_1$, and $ \beta_2$ . Thus, the convergence of  Algorithm 2 is proved.

\begin{balance}
\bibliography{IEEEabrv,IEEE_bib}
\end{balance}

\end{document}